\documentclass[aps,superscriptaddress,notitlepage,prxquantum,reprint]{revtex4-2}
\usepackage{epsfig,color}
\usepackage{graphicx}
\usepackage{dcolumn}
\usepackage{bm}
\usepackage{amsmath,amsfonts,amssymb,mathrsfs}
\usepackage{pstricks}
\usepackage{amsxtra}
\usepackage{cases}
\usepackage{amsthm}
\usepackage{braket}
\usepackage{natbib}
\usepackage{physics}
\usepackage{hyperref,color}
\usepackage{comment}
\definecolor{darkblue}{rgb}{0.0,0.0,0.7}
\hypersetup{colorlinks,breaklinks,linkcolor=darkblue,urlcolor=darkblue,anchorcolor=darkblue,citecolor=darkblue}

\providecommand{\ah}{{\hat{a}}}
\providecommand{\ad}{{\hat{a}^\dagger}}
\providecommand{\rhoh}{{\hat{\rho}}}
\providecommand{\shz}{{\hat{\sigma}_z}}

\providecommand{\Hh}{{\hat{H}}}

\providecommand{\Sh}[1]{\hat{S}^{(#1)}}
\providecommand{\Vh}[1]{\hat{V}^{(#1)}}
\providecommand{\VhNO}[3]{\hat{V}^{(#1)}_{#2#3}}
\providecommand{\ShNO}[3]{\hat{S}^{(#1)}_{#2#3}}

\begin{document}

\title{Driven-Dissipative Dynamics of Measurement-Induced State Transitions}

\author{Bo-Syun Pan}
\affiliation{Department of Physics and Center for Theoretical Physics, National Taiwan University, Taipei 106319, Taiwan}

\author{Yen-Hsiang Lin}
\affiliation{Department of Physics, National Tsing Hua University, Hsinchu 300013, Taiwan}
\affiliation{National Center for Excellence in Quantum Information Science and Engineering, National Tsing Hua University, Hsinchu 300013, Taiwan}

\author{Chiao-Hsuan Wang}
\email{chiaowang@phys.ntu.edu.tw}
\affiliation{Department of Physics and Center for Theoretical Physics, National Taiwan University, Taipei 106319, Taiwan}
\affiliation{Center for Quantum Science and Engineering, National Taiwan University, Taipei 106319, Taiwan}
\affiliation{Physics Division, National Center for Theoretical Sciences, Taipei 106319, Taiwan}

\begin{abstract}
Dispersive readout plays a central role in superconducting quantum computing, enabling quantum nondemolition measurements through a coupled microwave resonator. While stronger resonator drives can improve measurement speed and fidelity, they can also activate multi-photon resonances that trigger measurement-induced state transitions (MIST) out of the computational subspace. We develop a driven-dissipative framework for MIST that retains the quantum resonator response absent in semiclassical treatments and yields analytically derived transition-rate expressions. The upward and downward transition-rate profiles determine the steady-state populations and finite-time dynamics, enabling quantitative exploration across drive strength and detuning, while capturing the crossover between quantum-resolved and semiclassical-like dynamics. This framework reproduces the quantum dynamics and identifies a strongly driven regime beyond semiclassical Landau--Zener predictions, where delayed population transfer opens a finite-time readout window with high resonator photon population.  These results establish MIST as a predictive driven-dissipative process and characterize strongly driven regimes that can be leveraged for measurement optimization.
\end{abstract}

\maketitle

\section{Introduction\label{sec:introduction}}
Superconducting circuits have been established as one of the leading platforms for quantum computing hardware~\cite{Krantz2019,Kjaergaard2020}, where microwave resonators enable control and readout of artificial atoms in the circuit quantum electrodynamics architecture~\cite{Blais2021}. In particular, quantum nondemolition (QND) measurement of superconducting qubits can be implemented via dispersive readout, where the qubit induces a state-dependent frequency shift on a coupled microwave resonator, allowing state inference without directly perturbing the qubit~\cite{Blais2004,Wallraff2005,Walter2017}.
While increasing the drive strength generally improves readout contrast through increased resonator photon number, leakage into noncomputational qubit states under strong driving has been observed in both transmon and fluxonium superconducting-qubit platforms~\cite{Sank2016,Lescanne2019,Khezri2023,Bista2026,Wang2026,Fechant2025,Connolly2025,Zobrist2026,Dai2026}.
This phenomenon, commonly termed measurement-induced state transition (MIST)~\cite{Sank2016,Khezri2023,Nesterov2024,Singh2025,Chapple2026,Zobrist2026} or qubit ionization~\cite{Shillito2022,Dumas2024,Nojiri2024,Wang2026}, signals a breakdown of the dispersive regime and the associated QND behavior.

Multi-photon resonances between computational and higher excited qubit states have been identified as the primary mechanism behind such transitions~\cite{Sank2016}. These processes were recently captured by full quantum models that treat the exact qubit spectrum and the driven bosonic resonator beyond the dispersive approximation~\cite{Dumas2024}. However, simulating such models remains computationally demanding due to the large Hilbert space of the coupled nonlinear multilevel qubit-resonator system and the explicit time dependence of the drive. Moreover, although numerical simulations can capture the transition behavior quantitatively, they provide limited insight into the physical origin of the transition, making it difficult to identify the dominant mechanisms and the parameters that control its onset and dynamics.

To explore MIST behavior using tractable approximations, simplified quantum metrics such as qubit purity or dressed matrix elements have been proposed to qualitatively predict its onset~\cite{Nojiri2024,Nesterov2024}. For quantitative predictions, many theoretical and experimental analyses rely on semiclassical approaches that replace the resonator field with a coherent amplitude~\cite{Shillito2022,Khezri2023,Dumas2024,Wang2026}.  Despite their computational efficiency, these approaches miss the qubit-conditioned structure of the quantum resonator response, leading to discrepancies in the predicted MIST dynamics. Such limitations warrant an efficient framework that can capture both the quantitative behavior and the underlying physical mechanism of MIST beyond semiclassical approximations.

Here we present an analytically tractable and physically transparent description of MIST as a driven-dissipative nonequilibrium process.  By reducing the full qubit--resonator dynamics to the near-resonant multi-photon leakage subspace, we obtain upward and downward transition-rate profiles whose competition sets the relaxation timescale and finite-time dynamics. The imbalance between these rates determines the driven-dissipative dynamics of MIST, while its dependence on drive strength and detuning provides a quantitative framework for predicting how the applied drive controls the transition behavior.

By retaining the qubit-conditioned quantum resonator response, the reduced description reproduces the quantum dynamics and predicts where the transition departs from semiclassical Landau--Zener behavior. The distinguishability of the qubit-conditioned resonator states further characterizes the quantum-to-semiclassical crossover. Our analysis further reveals a finite-time readout window: for the representative parameters studied here, the resonator photon number reaches $\tilde n_{\mathrm{avg}}\gtrsim 25$ within $1~\mu\mathrm{s}$, well above the characteristic crossing photon number $n^{(g)}_{\rm cross}\simeq 6$, while the dressed-state population remains $\tilde P_g\gtrsim 0.95$. These findings offer analytical insight complementary to existing strategies for optimizing strongly driven dispersive readout~\cite{Bengtsson2024,Kurilovich2025,Mori2026}.

In the following, we first formulate MIST in the full qubit--resonator model and derive the reduced quantum model for the near-resonant multi-photon subspace in Sec.~\ref{sec:model}. In Sec.~\ref{sec:dynamics}, we develop the rate-equation description, showing how drive-dependent transition rates determine the relaxation dynamics and steady-state regimes. In Sec.~\ref{sec:beyond_semiclassical}, we analyze the quantum-to-semiclassical crossover by comparing the full quantum dynamics, reduced quantum description, and semiclassical model. We conclude in Sec.~\ref{sec:discussion} by discussing how these results elucidate the nonequilibrium dynamics underlying QND breakdown and provide an efficient analytical framework for optimizing dispersive readout.

\section{Quantum Model of MIST}\label{sec:model}
\subsection{Full Quantum Model}
\begin{figure}[htbp]
\begin{center}
\includegraphics[width=\linewidth]{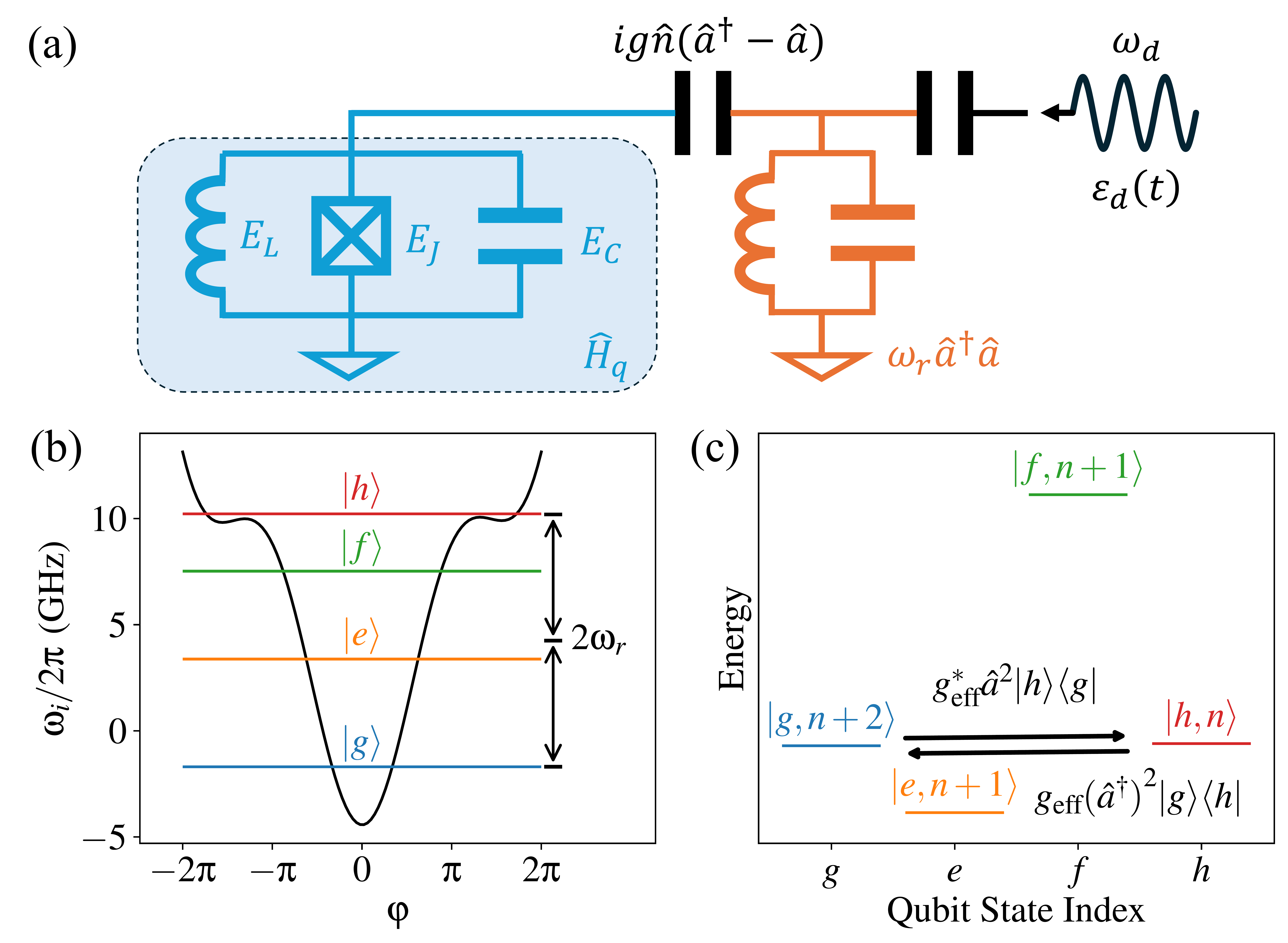}
\caption{(a) Schematic of dispersive qubit readout setup, where a fluxonium qubit is capacitively coupled to a microwave resonator driven by a classical field.
(b) Nonlinear energy spectrum of the fluxonium qubit. The transition between the ground and third excited states is near two-photon resonance with the readout resonator. (c) Effective two-photon transition between $\ket{g,n{+}2}$ and $\ket{h,n}$ in the reduced model, where $n$ is the resonator photon number. Other off-resonant states are neglected under the rotating wave approximation.}
    \label{fig:cQEDModel}
    \end{center}
\end{figure}

We begin with the full quantum model of a nonlinear superconducting qubit capacitively coupled to a driven microwave resonator~\cite{Dumas2024}, schematized in Fig.~\ref{fig:cQEDModel}(a). While the framework applies broadly to different superconducting-qubit platforms, such as transmons and fluxoniums, we focus on fluxonium as a concrete example with experimentally relevant parameters~\cite{Manucharyan2009,Nguyen2019}.
The qubit Hamiltonian is given by (taking $\hbar = 1$)
\begin{align}
    \hat{H}_q = 4E_C\hat{n}^2 + \frac{1}{2}E_L\hat{\varphi}^2 - E_J \cos(\hat{\varphi} - \varphi_{\rm ext}),
    \label{eqn:Hq}
\end{align}
where $\hat{\varphi}$, $\hat{n}$ are the normalized flux and charge operators of the fluxonium qubit, satisfying $\left[ \hat{\varphi}, \hat{n}\right] = i$. The parameters $E_C$, $E_L$, and $E_J$ denote the charging, inductive, and Josephson energies, respectively. The external flux is given by $\varphi_{\rm ext} = 2\pi \Phi_{\rm ext}/\Phi_0$, where $\Phi_0$ is the magnetic flux quantum. Numerical values of physical parameters used in this work are listed in Table~\ref{tab:FullParameters}. The qubit Hamiltonian can be diagonalized as $\hat{H}_q = \sum_{j} \omega_{j}\ketbra{j}{j}$, where $\ket{j}$ denotes the $j$-th qubit eigenstate, yielding a nonlinear energy spectrum as shown in Fig.~\ref{fig:cQEDModel}(b).

\begin{table}[htbp]
\caption{Fluxonium-resonator parameters used in the full quantum model. The circuit and resonator parameters are based on experimental device values, while the external flux $\varphi_{\rm ext}$ is chosen as a representative operating point near a well-isolated two-photon resonance between $\ket{g}$ and $\ket{h}\equiv\ket{3}$.}
\label{tab:FullParameters}
\centering
\renewcommand{\arraystretch}{1.1}
\begin{tabular}{l c l c}
\hline
Parameter & Value & Parameter & Value \\
\hline
$E_C/h$ & 4.43 GHz & $\omega_r/2\pi$ & 5.9436 GHz \\
$E_J/h$ & 0.795 GHz & $\omega_d/2\pi$ & 5.9436 GHz \\
$E_L/h$ & 0.89 GHz & $\kappa/2\pi$ & 4.086 MHz \\
$\varphi_{\rm ext}$ & 0.010 & $g/2\pi$ & 98 MHz \\
\hline
\end{tabular}
\end{table}

The full system Hamiltonian is $\Hh_{\rm full}(t) = \Hh_{qr} + \Hh_{d}(t)$, where the static qubit--resonator Hamiltonian reads
\begin{align}
    \hat{H}_{qr} = \omega_r\ad\ah + \hat{H}_q + ig\hat{n}(\ad - \ah).
    \label{eqn:Hqr}
\end{align}
Here, $\omega_r$ is the bare frequency of the resonator, and $i g \hat{n} (\ad - \ah)$ represents the capacitive coupling between the qubit and the resonator with strength $g$. The resonator is driven by a classical measurement field, modeled as a time-dependent drive Hamiltonian
\begin{align}
\hat{H}_d(t) = -2 i\varepsilon_d(t) \sin(\omega_d t)(\ad - \ah),
\label{eqn:Hdrive}
\end{align}
where $\omega_d$ and $\varepsilon_d$ are the drive frequency and the drive amplitude, respectively.

In the limit $\abs{g} \ll \abs{\Delta}$, with $\Delta \equiv \omega_q - \omega_r$ and $\omega_q \equiv \omega_{1} -\omega_{0}$, and assuming low photon occupation in the resonator, the system reduces to the standard dispersive coupling form when restricted to the dressed computational qubit subspace spanned by $\ket{g}\equiv\ket{0}$ and $\ket{e}\equiv\ket{1}$~\cite{Blais2004, Zhu2013},
\begin{align}
    \hat{H}_{qr} \approx\omega_r^{\prime}\ad\ah + 
    \omega_q^{\prime} \ketbra{e}{e} + \chi \ad\ah \ketbra{e}{e},
    \label{eqn:Hdispersive}
\end{align}
where $\omega_r^{\prime}$ and $\omega_q^{\prime}$ are the shifted frequencies and $\chi$ denotes the dispersive shift.  In this regime, the resonator acquires a qubit-state-dependent frequency shift, enabling quantum nondemolition (QND) measurement of the qubit state through the readout drive~\cite{Wallraff2005}.

At higher photon numbers, multi-photon processes can drive the qubit out of the computational subspace into higher excited states. For the parameters considered here, the dominant channel is a two-photon resonance between $\ket{g}$ and $\ket{h} \equiv \ket{3}$ at $2\omega_r \approx \omega_h - \omega_g$ [Fig.~\ref{fig:cQEDModel}(b)], which induces leakage out of the computational subspace and thereby breaks the QND readout character. However, the full quantum model is analytically intractable and numerically costly due to the large Hilbert space and time-dependent drive, posing challenges for understanding MIST dynamics. We therefore develop a reduced framework that captures the quantum resonator response and the driven-dissipative dynamics underlying MIST.

\subsection{Reduced Quantum Model}
When all qubit transitions remain far detuned from the resonator frequency,
$\abs{g\bra{i}\hat{n}\ket{j}}\ll\abs{\omega_j-\omega_i-\omega_r}$ for all $\ket{i},\ket{j}$,
we perform a Schrieffer--Wolff (SW) transformation together with the rotating-wave approximation to obtain an effective description restricted to the near-resonant subspace [Fig.~\ref{fig:cQEDModel}(c)]. 
The resulting Hamiltonian of the reduced quantum model reads (see Appendix~\ref{app:reduced_framework})
\begin{align}
    \hat{H}_{\rm eff} =& \delta_a \ad \ah + ( \delta_{g} + \chi_{g}\ad\ah)\ketbra{g}{g} + ( \delta_{h} + \chi_{h}\ad\ah)\ketbra{h}{h}\notag\\
    &+g_{\rm eff}{(\hat{a}^\dagger)}^2 \ketbra{g}{h}   + g_{\rm eff}^*\hat{a}^2 \ketbra{h}{g}  +\varepsilon_d (\ah + \ad),
    \label{eqn:Heff}
\end{align}
where $\delta_a=\omega_r-\omega_d$ is the resonator-drive detuning, $\delta_g$ and $\delta_h$ are the shifted energies of $\ket{g}$ and $\ket{h}$, $\chi_{g/h}$ are the dispersive couplings, and $g_{\rm eff}$ is the effective two-photon coupling strength. This effective Hamiltonian $\hat{H}_{\rm eff}$ is expressed in the dressed-state basis defined by the SW transformation and in the rotating frame. We further define $\delta_q\equiv\delta_h-\delta_g$ as the energy difference between the two qubit states. The derived parameters entering this reduced description are listed in Table~\ref{tab:ReducedParameters}.

\begin{table}[htbp]
\caption{Derived parameters entering the reduced quantum model. Unless otherwise specified, the reduced-model analysis uses the resonator linewidth $\kappa/2\pi = 4.086~\mathrm{MHz}$ and is evaluated at zero detuning, $\delta_a = 0$.}
\label{tab:ReducedParameters}
\centering
\renewcommand{\arraystretch}{1.1}
\begin{tabular}{l c l c}
\hline
Parameter & Value & Parameter & Value \\
\hline
$\delta_q/2\pi$ & 27.128 MHz & $\chi_g/2\pi$ & 4.029 MHz \\
$g_{\rm eff}/2\pi$ & 0.215 MHz & $\chi_h/2\pi$ & $-$ 0.739 MHz\\
\hline
\end{tabular}
\end{table}

The effective Hamiltonian resembles a two-photon Jaynes--Cummings interaction~\cite{Villas-Boas2019}, augmented by dispersive coupling terms and effective level shifts induced by the full qubit spectrum, and captures the relevant quantum dynamics.
The open-system dynamics within this reduced subspace, including resonator photon loss at rate $\kappa$, is governed by the Lindblad master equation
\begin{align}
    \partial_t \rhoh = -i[\hat{H}_{\rm eff}, \rhoh] + \kappa \mathcal{D} [\hat{a}](\rhoh),
    \label{eqn:MasterEffective}
\end{align}
where $\rhoh$ is the density matrix and $\mathcal{D}\left[\ah\right](\rhoh) \equiv \ah\rhoh\ad - \frac{1}{2}\{\ad\ah,  \rhoh\}$ is the Lindblad dissipator.

This reduced description highlights the two-photon qubit--resonator coupling responsible for measurement-induced transitions, providing analytical access to the qubit-conditioned quantum resonator response and nonequilibrium dynamics beyond semiclassical treatments. Defined in a reduced Hilbert space without explicit time dependence, our model enables analytical investigation of driven-dissipative dynamics and efficient numerical simulation at substantially lower cost compared to the full model. Although we focus on the two-photon resonance of fluxonium as a concrete example, the framework extends naturally to higher-order multi-photon resonances (see Appendix~\ref{app:multiphonon_extension}) and is broadly applicable across different superconducting qubit architectures.

\section{Driven-Dissipative Dynamics}\label{sec:dynamics}
Our analysis reveals that MIST is governed by driven-dissipative qubit dynamics. We first analyze the steady state of Eq.~\eqref{eqn:MasterEffective} by treating the two-photon coupling $g_{\rm eff}$ as a perturbation, $\hat{H}_{\rm eff} =\hat{H}^{(0)}+\hat{H}^{(1)}$, $ \hat{H}^{(1)}=g_{\rm eff}{(\hat{a}^\dagger)}^2 \ketbra{g}{h}   + g_{\rm eff}^*\hat{a}^2 \ketbra{h}{g}$. To leading order, the unperturbed Liouvillian $\mathcal{L}_0\rhoh = -i[\hat{H}^{(0)}, \rhoh] + \kappa\mathcal{D}[\ah](\rhoh)$ admits a degenerate steady-state manifold spanned by
\begin{align}
\rhoh_{ss}^{(0)} = \tilde{P}_g \ketbra{g}{g}\otimes \ketbra{\alpha_g}{\alpha_g} +\tilde{P}_h \ketbra{h}{h}\otimes \ketbra{\alpha_h}{\alpha_h},
\label{eqn:rhoss0}
\end{align}
with undetermined dressed-state populations $\tilde{P}_g$, $\tilde{P}_h$, where $\ket{\alpha_{g/h}}$ are coherent states conditioned on the qubit state, with amplitudes $\alpha_{g/h} = \frac{-\varepsilon_d}{\delta_a + \chi_{g/h} - i\kappa/2}$. The resonator adiabatically follows each qubit state and relaxes to the corresponding coherent state, while the two-photon coupling mediates transitions on a slower timescale.

The qubit-state populations evolve under the competition between upward and downward two-photon transitions. Including $\mathcal{L}_1\rhoh = -i[\Hh^{(1)}, \rhoh]$ perturbatively and applying adiabatic elimination in the weak coupling limit $\abs{g_{\rm eff}}\abs{\alpha_g}^2, \abs{g_{\rm eff}}\abs{\alpha_h}^2 \ll \kappa$~\cite{Gardiner2004}, we obtain the rate equation
\begin{align}
    \dot{\tilde{P}}_g = -\gamma_g \tilde{P}_g + \gamma_h\tilde{P}_h,
    \label{eqn:Pevolution}
\end{align}
where the drive-dependent transition rates $\gamma_g$, $\gamma_h$ can be analytically computed from the reduced quantum model (see Appendix~\ref{app:rate_equations}).
Enforcing $\tilde{P}_g + \tilde{P}_h = 1$, Eq.~\eqref{eqn:Pevolution} reduces to
\begin{align}
\dot{\tilde{P}}_g
=
-\gamma(\tilde{P}_g-\tilde{P}_g^{\rm ss}),
\label{eqn:Prelax}
\end{align}
where $\gamma=\gamma_g+\gamma_h$ and
$\tilde{P}_g^{\rm ss}=\gamma_h/(\gamma_g+\gamma_h)$.

For a fixed drive amplitude $\varepsilon_d$, the analytical solution of the rate equation and the associated average photon number are
\begin{align}
\tilde{P}_g(t)
&=
\tilde{P}_g^{\rm ss}
+
\left[
\tilde{P}_g(0)-\tilde{P}_g^{\rm ss}
\right]
e^{-\gamma t},
\notag\\
\tilde{P}_h(t)
&=
1-\tilde{P}_g(t),
\notag\\
\tilde n_{\rm avg}(t)
&=
\tilde{P}_g(t)|\alpha_g|^2
+
\tilde{P}_h(t)|\alpha_h|^2 .
\label{eqn:Pg_analytical_solution}
\end{align}
This describes an irreversible relaxation toward a unique steady state set by the imbalance between upward and downward two-photon transitions.

For fixed system parameters, the measurement drive determines the transition-rate imbalance and the corresponding driven-dissipative dynamics. In the zero-drive limit, $\varepsilon_d=0$, the resonator remains in the vacuum state and the upward transition vanishes, $\gamma_g=0$. The excited state can decay to the ground state through two-photon emission followed by resonator photon loss at a rate
\begin{align}
    \gamma_h = 4|g_{\rm eff}|^2\frac{\kappa}{[\delta_q - 2(\delta_a+\chi_g)]^2+\kappa^2},
    \label{eqn:ZeroDriveRateh}
\end{align}
which can be interpreted as a two-photon Purcell process. Here $\delta_q - 2(\delta_a+\chi_g)$ is the energy difference between $\ket{h,0}$ and $\ket{g,2}$, while the resonance is broadened by the two-photon decay rate $2\kappa$. The ground state is thus the unique steady state at zero drive. With the measurement drive on, both upward and downward transitions become drive dependent, creating a nonequilibrium interplay that sets the steady state and the relaxation timescale.

\begin{figure}[htbp]
\begin{center}
\includegraphics[width=\linewidth]{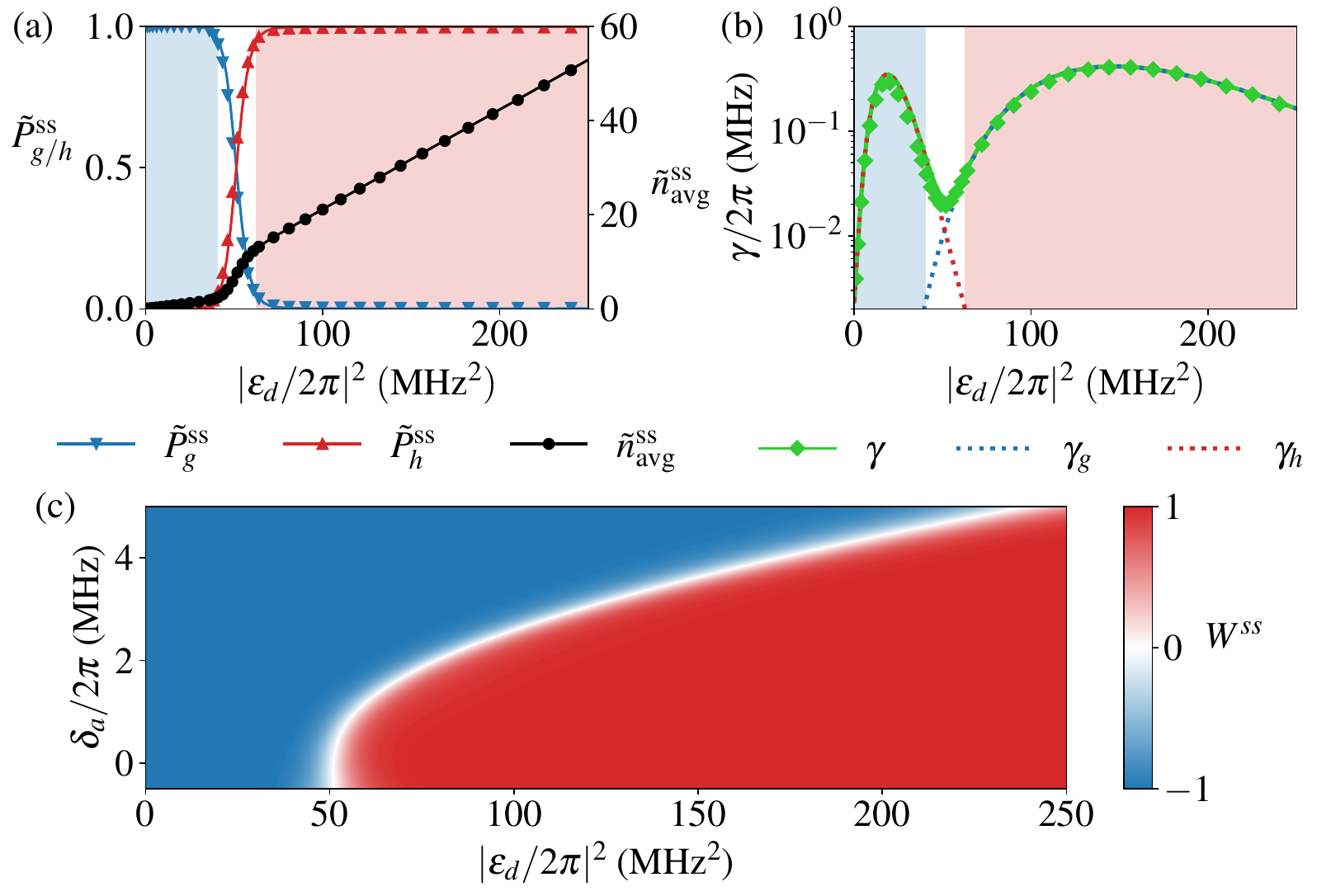}
\caption{(a) Steady-state qubit populations $\tilde{P}_{g}^{\rm ss}$ and $\tilde{P}_{h}^{\rm ss}$, and average photon number $\tilde{n}_{\mathrm{avg}}^{\rm ss}$  as functions of $\abs{\varepsilon_d}^2$.
(b) Relaxation rates $\gamma$, $\gamma_{g}$, and $\gamma_{h}$  as functions of $\abs{\varepsilon_d}^2$.
Markers in (a) and (b) denote numerical simulations of the reduced quantum model with resonator photon-number cutoff up to $n^{\rm max}=100$, while solid and dotted lines show analytical predictions from the rate equations.
(c) Steady-state inversion $W^{\rm ss} \equiv \tilde{P}_h^{\rm ss} - \tilde{P}_g^{\rm ss}$ in the parameter space spanned by $\abs{\varepsilon_d}^2$ and $\delta_a$, obtained from the analytical rate-equation solution.}
\label{fig:SteadyState}
\end{center}
\end{figure}

The drive amplitude controls the two-photon transitions through the qubit-state-dependent coherent-state amplitudes. For the upward transition $\ket{g,n}\rightarrow\ket{h,n-2}$, the branch-dependent resonance condition defines the $g$-branch crossing photon number
\begin{align}
n_{\rm cross}^{(g)}
=
\frac{\delta_q-2(\delta_a+\chi_h)}{\chi_g-\chi_h}.
\end{align}
The upward rate $\gamma_g$ is maximized when the $g$-conditioned mean photon number $\bar n_g\equiv|\alpha_g|^2$ approaches $n_{\rm cross}^{(g)}$, while the downward rate $\gamma_h$ is maximized when $\bar n_h\equiv|\alpha_h|^2$ approaches $n_{\rm cross}^{(h)}=n_{\rm cross}^{(g)}-2$. For the set of parameters in Table~\ref{tab:ReducedParameters} at $\delta_a=0$, these conditions give $n_{\rm cross}^{(g)}\simeq 6$ and $n_{\rm cross}^{(h)}\simeq 4$.  Since $\bar n_g$ and $\bar n_h$ scale differently with $|\varepsilon_d|^2$, the upward and downward transition rates acquire distinct drive dependences. The resulting transition-rate profiles and steady-state populations, shown in Fig.~\ref{fig:SteadyState}(a),(b), agree well with numerical simulations of the reduced quantum model.

The distinct transition-rate profiles give rise directly to a three-stage structure in MIST, as identified in Fig.~\ref{fig:SteadyState}(a),(b). Under weak drives, $\gamma_h \gg \gamma_g$, so the qubit remains predominantly in the ground state and the measurement is approximately QND; this defines the \textbf{sub-MIST} regime (blue-shaded region). As the drive increases, $\gamma_g$ and $\gamma_h$ become comparable, leading to substantial population redistribution between $\ket{g}$ and $\ket{h}$ and defining a finite \textbf{MIST window}, which we identify operationally by $5\% < \tilde{P}^{\rm ss}_h < 95\%$ (white window). At stronger drives, the system moves beyond this window into the \textbf{super-MIST} regime (red-shaded region). The same three-stage structure extends over the parameter space spanned by the drive amplitude $\varepsilon_d$ and detuning $\delta_a$, as visualized by the steady-state inversion $W^{\rm ss} \equiv \tilde{P}_h^{\rm ss} - \tilde{P}_g^{\rm ss}$ in Fig.~\ref{fig:SteadyState}(c).

In the present quantum treatment, the resonator response is determined by distinct qubit-conditioned coherent states and is therefore intrinsically branch dependent. The steady-state photon number in the dressed basis,
\begin{align}
\tilde{n}_{\mathrm{avg}}^{\rm ss}
=
\tilde{P}_g^{\rm ss} |\alpha_g|^2+\tilde{P}_h^{\rm ss} |\alpha_h|^2,
\end{align}
follows the same three-stage structure. In the sub-MIST regime, $\tilde n_{\mathrm{avg}}^{\rm ss}\approx \bar n_g$ and increases approximately linearly with $|\varepsilon_d|^2$. Across the MIST window, population transfer from the $g$ branch to the $h$ branch causes a rapid increase in $\tilde n_{\mathrm{avg}}^{\rm ss}$. Beyond this window, $\tilde n_{\mathrm{avg}}^{\rm ss}\approx \bar n_h$, leading to an approximately linear growth with a larger slope. While the resonant photon numbers still indicate the peak locations of the two transition-rate profiles, their competition extends the steady-state response into a finite MIST window. The drive amplitude $\varepsilon_d$ therefore serves as the natural control parameter for the driven-dissipative dynamics.

These results establish MIST as a driven-dissipative nonequilibrium process controlled by readout-drive-dependent transition rates. Within the reduced quantum model, branch-dependent resonance conditions determine the upward and downward two-photon transition-rate profiles. Their imbalance gives rise to the observed three-stage steady-state structure and resonator response, and determines the time evolution analytically. This framework provides a quantitatively accurate, physically transparent, and computationally efficient description of MIST, making its steady-state structure and relaxation dynamics accessible without demanding long-time simulations of the full quantum model.

\section{Beyond Semiclassical Predictions}\label{sec:beyond_semiclassical}
The driven-dissipative framework reveals dynamical features of MIST that are not captured by semiclassical and Landau--Zener descriptions.  In semiclassical approaches, the resonator field is treated as a coherent amplitude, and the transition is interpreted as a diabatic process driven by an effective classical field~\cite{Shillito2022,Dumas2024,Wang2026}. In contrast, the quantum description developed here retains the quantized resonator mode, which gives rise to transition-rate imbalance and nonequilibrium dynamics. In the following, we examine the crossover between quantum and semiclassical behavior and benchmark these descriptions using the full quantum model.

\subsection{Quantum-to-Semiclassical Crossover}

The key distinction between the quantum and semiclassical descriptions of MIST lies in the treatment of the resonator during the transition.  In a semiclassical description, the resonator enters as a classical field described by a single evolving complex amplitude. In the quantum description, the resonator is instead represented by qubit-conditioned states, and their separation remains explicit in the transition dynamics. We characterize the separation between the two conditional resonator states by the distinguishability $D \equiv \abs{\alpha_g-\alpha_h}^2$, which measures their distance in phase space~\cite{Gambetta2006}.

Near the upward two-photon transition, the distinguishability between the qubit-state-dependent resonator states is characterized by
\begin{align}
D^{(g)}_{\rm cross}
\equiv
D\Big|_{\bar n_g=n_{\rm cross}^{(g)}}
=
n_{\rm cross}^{(g)}
\frac{(\chi_g-\chi_h)^2}
{(\delta_a+\chi_h)^2+\kappa^2/4}.
\end{align}
This quantity separates the transition dynamics into two qualitatively different limits.  When $D^{(g)}_{\rm cross} \gg 1$, the two conditional coherent states are well separated, and the transition is distributed over multiple photon-number-resolved channels. When $D^{(g)}_{\rm cross} \ll 1$, the two conditional states strongly overlap, so the resonator can be approximated by a single coherent-state response and the dynamics approaches an unresolved, more semiclassical-like behavior.

\begin{figure}[t]
\centering
\includegraphics[width=0.85\linewidth]{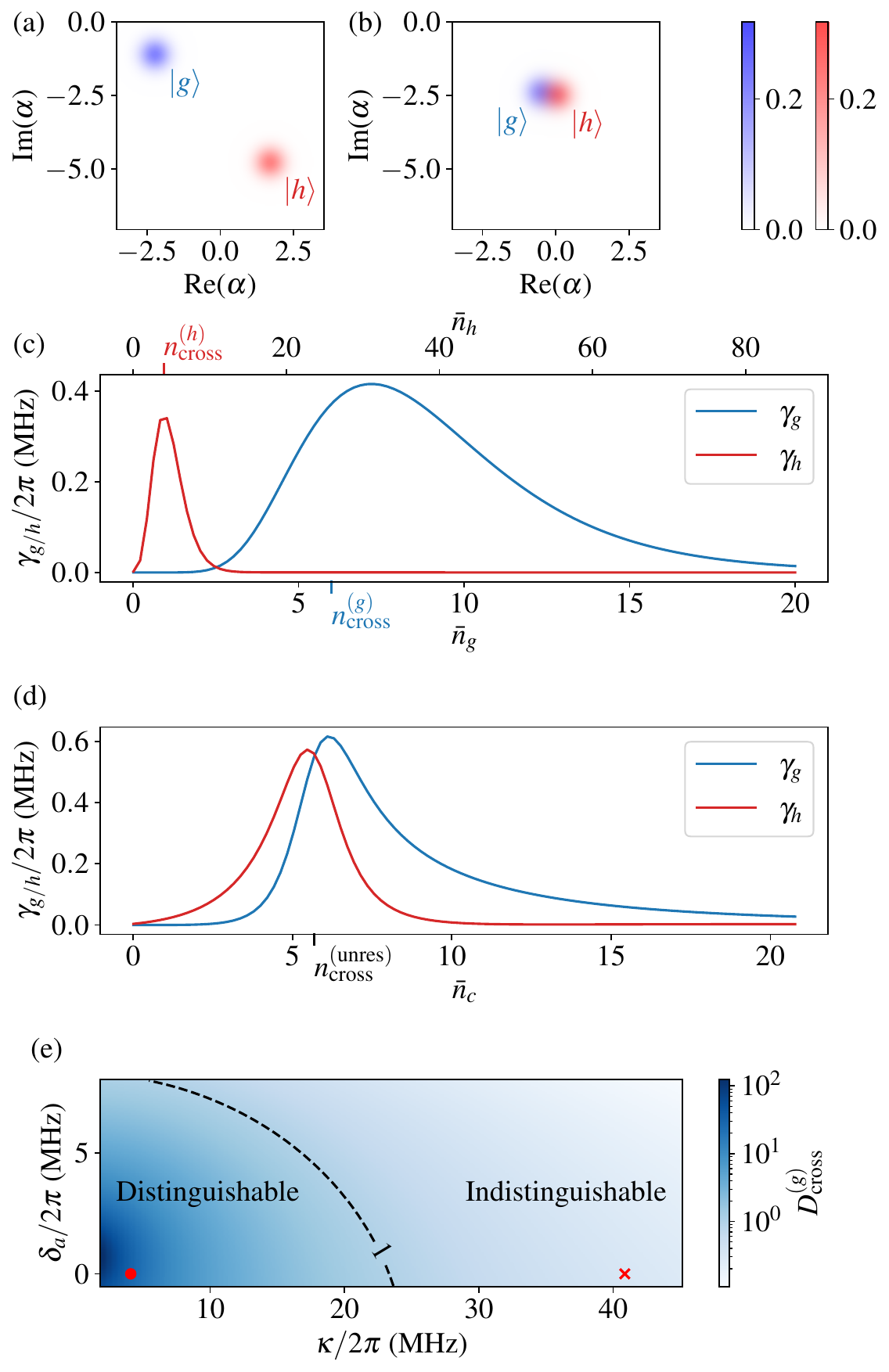}
\caption{
Distinguishability and transition profiles.
(a) Wigner functions of the conditional displaced states at the ground-state resonance $\bar n_g=n_{\rm cross}^{(g)}$, evaluated for the reduced-model parameters listed in Table~\ref{tab:ReducedParameters}.
(b) Wigner functions of the conditional displaced states at the same resonance point for a representative strong-dissipation case with $\kappa$ increased by a factor of ten.
(c) Transition-rate profiles for the manuscript parameters in the resolved regime, shown as functions of $\bar n_g$ and $\bar n_h$.
(d) Transition profile for the representative strong-dissipation case in the unresolved regime, shown as a function of $\bar{n}_c \equiv |\alpha_c|^2 = |\varepsilon_d|^2/(\delta_a^2+\kappa^2/4)$.
(e) Near-resonant phase-space separation $D^{(g)}_{\rm cross}$ as a function of $\kappa$ and $\delta_a$, with all other parameters fixed to the values listed in Table~\ref{tab:ReducedParameters}. The circle and cross mark the representative resolved and unresolved parameter sets used throughout the manuscript.}
\label{fig:Distinguishability}
\end{figure}

Readout-relevant parameters typically lie in the resolved regime $D^{(g)}_{\rm cross} \gg 1$, since useful dispersive readout requires a clear qubit-state-dependent resonator response and a non-negligible crossing photon number. For the parameters considered in this manuscript, the system lies in the resolved regime near the two-photon resonance, consistent with the well-separated Wigner functions shown in Fig.~\ref{fig:Distinguishability}(a). The transition-rate profiles can then be approximated by Gaussian-like envelopes,
\begin{align}
\gamma_{g/h} \propto \exp\!\left[-\frac{\bigl(\bar n_{g/h}-n_{\rm cross}^{(g/h)}\bigr)^2}{2n_{\rm cross}^{(g/h)}}\right],
\end{align}
with maxima at $\bar n_{g/h}\approx n_{\rm cross}^{(g/h)}$ and widths scaling as $\sqrt{n_{\rm cross}^{(g/h)}}$ (see Appendix~\ref{app:regimes} for details), consistent with the branch-resolved profiles shown in Fig.~\ref{fig:Distinguishability}(c).

When the two conditional resonator states strongly overlap near the two-photon resonance, the dynamics enters the unresolved limit, as shown in Fig.~\ref{fig:Distinguishability}(b). This limit typically requires either strong resonator dissipation or large resonator-drive detuning, and is therefore not the usual operational regime for dispersive qubit readout. In this case, the branch-resolved transition profiles approach a common Lorentzian line shape,
\begin{align}
\gamma
\propto
\frac{\Gamma_\phi}{
\left[
\delta_q-(\chi_g-\chi_h)\bar n_c
\right]^2
+
\Gamma_\phi^2},
\end{align}
where $\alpha_c \equiv -\varepsilon_d/(\delta_a-i\kappa/2)$, $\bar n_c\equiv|\alpha_c|^2$, and $\Gamma_\phi\equiv \kappa D/2$ is the measurement-induced dephasing rate~\cite{Gambetta2006}. The Lorentzian is centered at the unresolved crossing photon number $n_{\rm cross}^{\rm (unres)}\equiv\delta_q/(\chi_g-\chi_h)$ (see Appendix~\ref{app:regimes} for details). The resulting transition profiles are shown in Fig.~\ref{fig:Distinguishability}(d), where they exhibit a semiclassical-like sharp transition near the common resonance.

The near-resonant phase-space separation therefore provides a direct metric for the quantum-to-semiclassical crossover in MIST.  Figure~\ref{fig:Distinguishability}(e) summarizes the distinguishability $D^{(g)}_{\rm cross}$ in the parameter space of the resonator dissipation $\kappa$ and the drive detuning from the resonator mode, $\delta_a$. For readout-relevant parameters, the system lies in the resolved regime, where the qubit-conditioned pointer states are well separated and the transition is distributed over branch-resolved channels. In the opposite limit of strong resonator dissipation or large drive detuning, this phase-space separation is suppressed and the driven-dissipative dynamics approaches a more semiclassical behavior.

\subsection{Time Evolution and Transient Regimes}
The distinction between the quantum and semiclassical descriptions becomes even more apparent in the time evolution. To characterize these dynamical differences, we use the full quantum model as a benchmark and compare its predictions with those of the reduced quantum model, the analytical rate-equation solution, and the semiclassical model with backaction~\cite{Wang2026}. Starting from the system ground state, we first analyze the time-dependent qubit populations and mean photon number in both the resolved and unresolved cases. We then focus on the readout-relevant resolved regime and examine fixed-time snapshots over a range of drive strengths to identify distinct readout features.

\begin{figure}[htbp]
\begin{center}
\includegraphics[width=\linewidth]{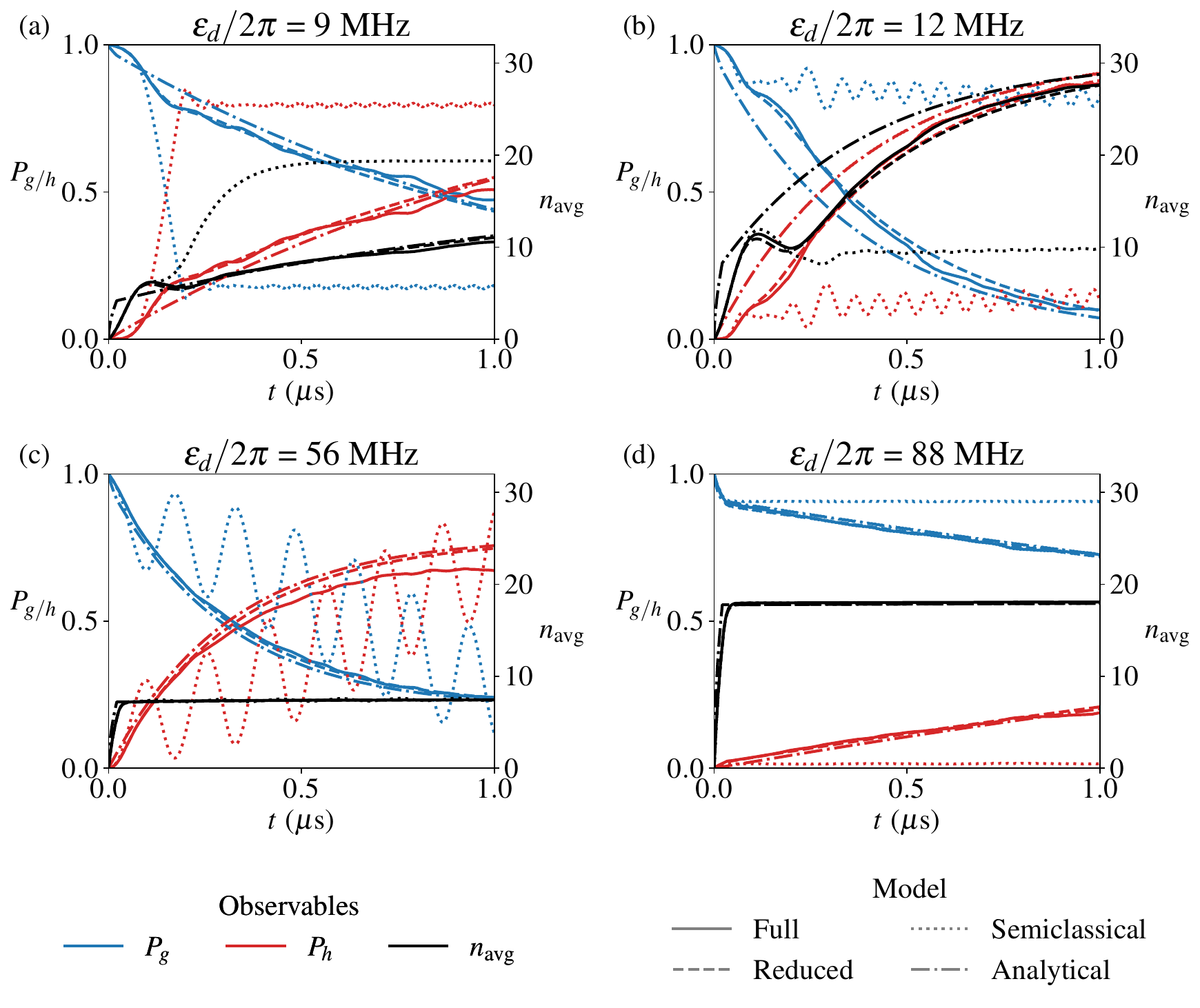}
\caption{
Time evolution of $P_g$, $P_h$, and $n_{\mathrm{avg}}$ in the bare-state basis. Panels (a) and (b) correspond to the resolved case, while panels (c) and (d) correspond to the unresolved case.
Panels (a) and (c) are chosen near the transition predicted by the semiclassical model, whereas panels (b) and (d) are chosen at a stronger drive.
Starting from the ground state of the qubit and resonator, numerical simulations are performed using QuTiP~\cite{Lambert2026} with a qubit level cutoff $j^{\rm max}=4$ for the full and semiclassical models, and a resonator photon-number cutoff up to $n^{\rm max}=100$ for the full and reduced quantum models.
The analytical curves follow the rate equation in Eq.~\eqref{eqn:Pg_analytical_solution}.}
\label{fig:EvolutionComparison}
\end{center}
\end{figure}

A direct comparison of the time evolution under representative drive conditions reveals a clear dynamical mismatch between the quantum and semiclassical descriptions. For this comparison, the reduced-model and rate-equation results are transformed back to the bare-state observables used in the full quantum and semiclassical simulations. In the resolved case, at a drive strength near the main transition predicted by the semiclassical model [Fig.~\ref{fig:EvolutionComparison}(a)], the quantum and semiclassical descriptions exhibit different transition timescales and population dynamics. Within the quantum description, the analytical rate-equation solution follows the full and reduced quantum dynamics well, apart from the short-time deviation expected from the adiabatic elimination of the resonator response. Once the qubit-conditioned resonator response is established, the subsequent population dynamics is governed by the driven-dissipative rate imbalance.

In the resolved case at a stronger drive deep in the super-MIST regime [Fig.~\ref{fig:EvolutionComparison}(b)], the mismatch becomes more pronounced. The quantum models and the rate-equation solution continue to show population transfer into $\ket{h}$, whereas the semiclassical model predicts that the qubit remains largely in the ground state. This semiclassical prediction follows the Landau--Zener picture of an increasingly diabatic passage at stronger drive~\cite{Shillito2022,Dumas2024,Wang2026}. The quantum models and the rate-equation solution reveal a persistent driven-dissipative population transfer in the strongly driven regime, a feature missed by the semiclassical Landau--Zener picture.

To connect this time-domain comparison with the crossover discussed in Fig.~\ref{fig:Distinguishability}, we also examine the unresolved regime in Fig.~\ref{fig:EvolutionComparison}(c),(d). Here the two qubit-conditioned resonator states strongly overlap, so the overall dynamics becomes more semiclassical-like. Near the semiclassical MIST region [Fig.~\ref{fig:EvolutionComparison}(c)], quantum and semiclassical models show closer agreement than in the resolved case. At stronger drive [Fig.~\ref{fig:EvolutionComparison}(d)], however, the semiclassical model still misses the population transfer into $\ket{h}$ predicted by the quantum models and the rate-equation solution. Figure~\ref{fig:EvolutionComparison} therefore supports the driven-dissipative description of MIST: the analytical rate-equation dynamics captures the quantum population transfer, while the semiclassical description shows quantitative and qualitative deviations, especially in the strongly driven regime.

\begin{figure}[htbp]
\begin{center}
\includegraphics[width=\linewidth]{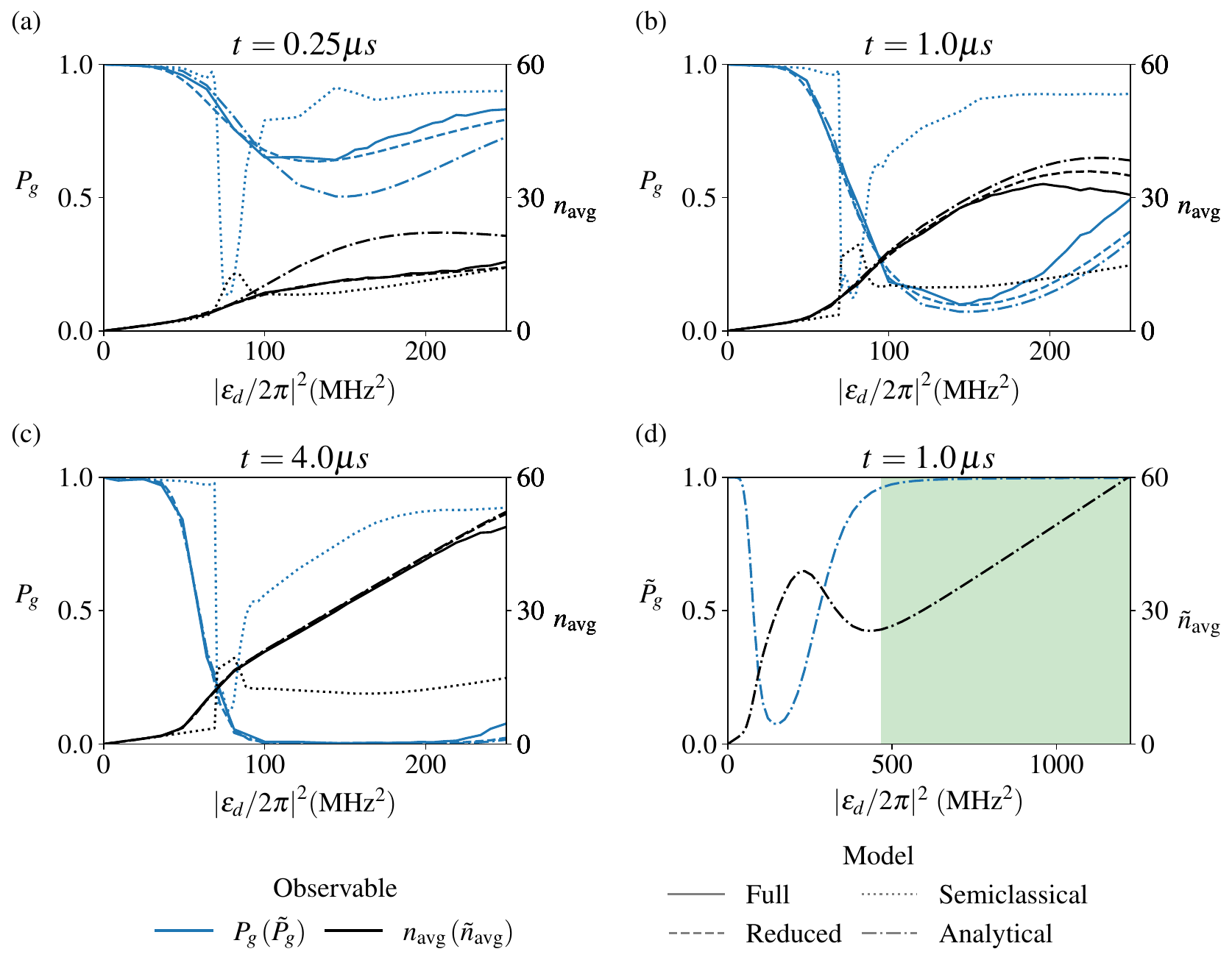}
\caption{
Fixed-time qubit population and average photon number across drive strengths.
(a--c) Snapshots of the bare-basis ground-state population $P_g$ and photon number $n_{\rm avg}$ at $t=0.25$, $1.0$, and $4.0~\mu\mathrm{s}$, comparing the full quantum model, the reduced quantum model, the analytical rate-equation solution, and the semiclassical model with backaction.
Numerical simulations are performed using QuTiP~\cite{Lambert2026}, with a qubit-level cutoff $j^{\rm max}=4$ for the full and semiclassical models and a photon-number cutoff $n^{\rm max}=100$ for the full and reduced models.
(d) Analytical rate-equation prediction for the dressed-basis ground-state population $\tilde P_g$ and average photon number $\tilde n_{\rm avg}$ at $t=1~\mu\mathrm{s}$ over an extended drive range. The green region marks the finite-time readout window where $\tilde P_g>95\%$ and $\tilde n_{\rm avg}$ is large.
}
\label{fig:FixedTime}
\end{center}
\end{figure}

We next examine fixed-time snapshots as a function of drive strength in the resolved case. At fixed times, the semiclassical model shows a narrow drive window for population transfer followed by a ground-state revival at stronger drive, consistent with the Landau--Zener picture. At $t = 0.25~\mu\mathrm{s}$ [Fig.~\ref{fig:FixedTime}(a)], the quantum models show only weak population transfer near the intermediate-drive region. By $t = 1~\mu\mathrm{s}$ [Fig.~\ref{fig:FixedTime}(b)], the difference becomes more apparent, with population transfer over a broader drive range than the semiclassical prediction. At $t = 4~\mu\mathrm{s}$ [Fig.~\ref{fig:FixedTime}(c)], this separation is most pronounced in the strongly driven regime, where the quantum models show clear escape from the ground state, whereas the semiclassical result remains largely frozen.

The fixed-time snapshots show that, at an experimentally relevant readout time of $t=1~\mu\mathrm{s}$, the strongly driven regime can retain a large ground-state population over a broad range of drive strengths. This short-time behavior may resemble Landau--Zener-like suppression, but at longer times the quantum models develop driven-dissipative population transfer into $\ket{h}$. This delayed leakage, arising from the slow driven-dissipative relaxation, opens a finite-time readout window before the measurement-induced transition becomes substantial.

We identify this high-drive readout window in the green region of Fig.~\ref{fig:FixedTime}(d). In this window, the qubit remains predominantly in the dressed ground-state component ($\tilde P_g>95\%$) at $t=1~\mu\mathrm{s}$ while the resonator becomes highly populated. This regime lies beyond the semiclassical Landau--Zener picture, which associates strong drive with transition suppression and limited resonator excitation ($n_{\mathrm{avg}}\lesssim 15$).  Instead, the quantum driven-dissipative dynamics reveals a high-photon-number readout condition before appreciable qubit leakage develops. 
These finite-time dynamics connect the microscopic transition-rate imbalance to the operational limits and opportunities of strongly driven dispersive readout.

\section{Discussion}\label{sec:discussion}

In summary, we have developed a driven-dissipative nonequilibrium framework for measurement-induced state transition. It yields analytical predictions for the steady state, relaxation timescale, and finite-time dynamics without relying on costly long-time simulations of the full system. The framework applies to MIST processes driven by multi-photon couplings, including higher-order resonances and other superconducting qubit architectures. By retaining the qubit-conditioned resonator states and the resulting transition-rate imbalance, this approach explains the quantum-to-semiclassical crossover, captures strong-drive behavior beyond semiclassical treatments, and uncovers a transient regime favorable for fast high-contrast readout.

The reduced quantum model provides a compact description when the MIST dynamics is dominated by an isolated near-resonant multi-photon transition between two qubit levels. Quantitative deviations from the full quantum model can arise under excessive drive power or resonator loss, where additional multi-photon resonances may become relevant and the single-transition perturbative treatment may become insufficient. Future extensions may retain more qubit levels and include higher-order corrections, such as Kerr nonlinearities and induced dissipative processes, to describe richer MIST dynamics beyond the two-level, single-transition description.

Looking ahead, the driven-dissipative rate framework developed here enables tractable exploration of transition rates across the parameter space of drive amplitude and detuning, making time-dependent pulse shaping and dynamically varied drive detunings a natural next step for further improving readout performance. More broadly, this framework may be extended to other drive-induced unwanted state transitions, including parasitic transitions~\cite{Singh2025}, material- or defect-induced effects~\cite{Bista2026}, and mechanisms beyond intrinsic multi-photon transitions~\cite{Dai2026}. The same approach may also help analyze higher-order multi-photon resonances in strongly coupled settings, such as multi-photon Jaynes--Cummings models~\cite{Carmichael2015,Li2024}.
These results establish an efficient and predictive framework for optimizing measurement protocols in strongly driven superconducting systems by revealing the nonequilibrium dynamics underlying QND breakdown.

\begin{acknowledgments}
The authors thank Huan-Hsuan Kung for helpful discussions. C.-H.W. acknowledges funding support from the National Science and Technology Council, Taiwan, under Grants No.~111-2112-M-002-049-MY3, No.~114-2119-M-007-013, No.~114-2124-M-002-003, and No.~114-2112-M-002-021-MY3, and from National Taiwan University under Grants No.~114L895001 and No.~115L893701.  C.-H.W. is also grateful for support from the National Taiwan University Eminence Scholar Fellowship and from the Physics Division, National Center for Theoretical Sciences, Taiwan. Y.-H.L. is supported by the National Science and Technology Council in Taiwan under Grants No.~113-2119-M-007-008 and No.~114-2119-M-007-011, by the Taiwan Centers of Excellence from the Ministry of Education, and by the Center for Quantum Technology, Project funded by the Ministry of Education in Taiwan.
\end{acknowledgments}

\appendix

\section{Derivation of the Reduced Quantum Model}
\label{app:reduced_framework}

Starting from the full quantum model in Eqs.~\eqref{eqn:Hqr} and~\eqref{eqn:Hdrive}, we derive the reduced Hamiltonian in Eq.~\eqref{eqn:Heff} for the near-resonant subspace shown in Fig.~\ref{fig:cQEDModel}(c). We assume that all single-photon qubit-resonator transitions remain far detuned, $\abs{g\bra{i}\hat{n}\ket{j}} \ll \abs{\omega_j-\omega_i-\omega_r}$ for all qubit states $\ket{i}$ and $\ket{j}$, while the dominant near-resonant process is the two-photon transition between $\ket{g}$ and $\ket{h}$, $\omega_h-\omega_g \approx 2\omega_r$.

We first decompose the static qubit-resonator Hamiltonian as
\begin{align}
    \Hh_{qr}
    =
    \Hh_0+\Vh{1},
\end{align}
with
\begin{align}
    \Hh_0
    =
    \Hh_q+\omega_r\ad\ah,
    \qquad
    \Vh{1}
    =
    ig\hat{n}(\ad-\ah).
\end{align}
Performing a Schrieffer--Wolff transformation~\cite{Schrieffer1966} perturbatively eliminates the off-resonant qubit-resonator coupling and yields the effective interaction in the near-resonant subspace,
\begin{align}
    \Hh_{qr}\rightarrow e^{\Sh{1}}\Hh_{qr}e^{-\Sh{1}},
\end{align}
with the first-order generator
\begin{align}
    \Sh{1}
    =
    \ShNO{1}{1}{0}\ad+\ShNO{1}{0}{1}\ah,
\end{align}
where $\bra{i}\ShNO{1}{1}{0}\ket{j} = -{ig\bra{i}\hat{n}\ket{j}}/{(\omega_{ij}-\omega_r)}$, $\ShNO{1}{0}{1}=-(\ShNO{1}{1}{0})^\dagger$, and $\omega_{ij}\equiv\omega_j-\omega_i$. Expanding the transformed Hamiltonian to second order gives
\begin{align}
    e^{\Sh{1}}\Hh_{qr}e^{-\Sh{1}}
    =
    \Hh_0+\Vh{2}+\mathcal{O}(g^3),
\end{align}
where
\begin{align}
    \Vh{2}
    & =
    \frac{1}{2}[\Sh{1},\Vh{1}]\notag\\
    & =
    \VhNO{2}{1}{1}\ad\ah
    +
    \VhNO{2}{0}{0}
    +
    \VhNO{2}{2}{0}(\ad)^2
    +
    \VhNO{2}{0}{2}\ah^2.
\end{align}
Restricting the qubit Hilbert space to the near-resonant subspace $\{\ket{g},\ket{h}\}$ and applying the rotating-wave approximation, we retain only the diagonal components of $\VhNO{2}{1}{1}$ and $\VhNO{2}{0}{0}$ and the resonant matrix element of $\VhNO{2}{2}{0}$. The transformed static Hamiltonian then becomes
\begin{align}
    e^{\Sh{1}}\Hh_{qr}e^{-\Sh{1}}
    &\approx
    \omega_r\ad\ah
    +
    (\omega_g+\Lambda_g+\chi_g\ad\ah)\ketbra{g}{g}   \notag\\
     &\quad+
    (\omega_h+\Lambda_h+\chi_h\ad\ah)\ketbra{h}{h}
    \notag\\
    &\quad
    +
    g_{\rm eff}(\ad)^2\ketbra{g}{h}
    +
    g_{\rm eff}^*\ah^2\ketbra{h}{g},
\end{align}
where $\chi_i=\bra{i}\VhNO{2}{1}{1}\ket{i}$ and $\Lambda_i=\bra{i}\VhNO{2}{0}{0}\ket{i}$ denote the dispersive and energy shifts, and $g_{\rm eff}=\bra{g}\VhNO{2}{2}{0}\ket{h}$ is the effective two-photon coupling. The explicit coefficient expressions are given in the Supplemental Material~\cite{note1}.

Applying the same transformation to the drive Hamiltonian in Eq.~\eqref{eqn:Hdrive} gives, to leading order,
\begin{align}
    e^{\Sh{1}}\Hh_d e^{-\Sh{1}}
    =
    \Hh_d+[\Sh{1},\Hh_d]+\mathcal{O}(\varepsilon_d g^2).
\end{align}
Under the same far-detuned condition, the commutator term $[\Sh{1},\Hh_d]$ is off resonant and is neglected within the rotating-wave approximation.  The transformed drive therefore reduces to
\begin{align}
    e^{\Sh{1}}\Hh_d e^{-\Sh{1}}
    \approx
    \varepsilon_d \ad e^{-i\omega_d t}
    +
    \varepsilon_d \ah e^{i\omega_d t}.
\end{align}

Applying the same transformation to the dissipator and retaining the leading-order terms gives the same Lindblad form within the rotating-wave approximation, so that the reduced dynamics is
\begin{align}
    \partial_t \rhoh
    =
    -i[\Hh_{\rm red},\rhoh]
    +
    \kappa \mathcal{D}[\ah](\rhoh),
\end{align}
with
\begin{align}
    \Hh_{\rm red}
    &\approx
    \omega_r\ad\ah
    +
    (\omega_g+\Lambda_g+\chi_g\ad\ah)\ketbra{g}{g}    \notag\\
    &\quad
    +
    (\omega_h+\Lambda_h+\chi_h\ad\ah)\ketbra{h}{h}
    \notag\\
    &\quad
    +
    g_{\rm eff}(\ad)^2\ketbra{g}{h}
    +
    g_{\rm eff}^*\ah^2\ketbra{h}{g}    \notag\\
    &\quad
    +
    \varepsilon_d\left(\ad e^{-i\omega_d t}+\ah e^{i\omega_d t}\right).
\end{align}

Finally, moving to the rotating frame generated by
\begin{align}
    \hat{U}_r(t)
    =
    \exp\left[
    -i\left(\omega_d\ad\ah+2\omega_d\ketbra{h}{h}\right)t
    \right],
\end{align}
we obtain the time-independent effective Hamiltonian quoted in Eq.~\eqref{eqn:Heff},
\begin{align}
    \Hh_{\rm eff}
    =&\,
    \delta_a \ad\ah
    +
    (\delta_g+\chi_g\ad\ah)\ketbra{g}{g}
    +
    (\delta_h+\chi_h\ad\ah)\ketbra{h}{h}
    \notag\\
    &+
    g_{\rm eff}(\ad)^2\ketbra{g}{h}
    +
    g_{\rm eff}^*\ah^2\ketbra{h}{g}
    +
    \varepsilon_d(\ah+\ad),
\end{align}
where $\delta_a=\omega_r-\omega_d$, $\delta_g=\omega_g+\Lambda_g$, and $\delta_h=\omega_h+\Lambda_h-2\omega_d$. The reduced dynamics therefore takes the form of Eq.~\eqref{eqn:MasterEffective}. Additional derivation details are provided in the Supplemental Material~\cite{note1}.

\section{Driven-Dissipative Population Dynamics and Rate Equations}
\label{app:rate_equations}

Starting from Eq.~\eqref{eqn:MasterEffective}, we treat the two-photon coupling perturbatively as in Sec.~III and write
\begin{align}
    \Hh_{\rm eff}
    =
    \Hh^{(0)}+\Hh^{(1)},
\end{align}
with
\begin{align}
    \Hh^{(1)}
    =
    g_{\rm eff}(\ad)^2\ketbra{g}{h}
    +
    g_{\rm eff}^*\ah^2\ketbra{h}{g}.
\end{align}
The corresponding Liouvillian is decomposed as
\begin{align}
    \dot{\rhoh}
    =
    \mathcal{L}_0\rhoh+\mathcal{L}_1\rhoh,
\end{align}
where
\begin{align}
    \mathcal{L}_0\rhoh
    &=
    -i[\Hh^{(0)},\rhoh]+\kappa\mathcal{D}[\ah](\rhoh),
    \\
    \mathcal{L}_1\rhoh
    &=
    -i[\Hh^{(1)},\rhoh].
\end{align}

To treat the zeroth-order dynamics, we move to the conditional displaced frame
\begin{align}
    \check{\rho}
    =
    \hat{D}_c^\dagger \rhoh \hat{D}_c,
    \qquad
    \hat{D}_c
    =
    \ketbra{g}{g}\hat{D}(\alpha_g)+\ketbra{h}{h}\hat{D}(\alpha_h),
\end{align}
where $\hat{D}(\alpha)=\exp(\alpha \ad-\alpha^* \ah)$ and
\begin{align}
    \alpha_{g/h}
    =
    -\frac{\varepsilon_d}{\delta_a+\chi_{g/h}-i\kappa/2}.
\end{align}
Defining
\begin{align}
    \hat{\alpha}
    &=
    \alpha_g\ketbra{g}{g}
    +
    \alpha_h\ketbra{h}{h},
    \\
    \check{\chi}
    &=
    (\delta_a+\chi_g)\ketbra{g}{g}
    +
    (\delta_a+\chi_h)\ketbra{h}{h},
\end{align}
the zeroth-order master equation in the transformed frame reads, dropping an irrelevant constant,
\begin{align}
    \check{\mathcal{L}}_0\check{\rho}
    =&
    -i\left[\check{\chi}\ad\ah+\frac{1}{2}\omega^{D}_q\shz,\check{\rho}\right]
    +\kappa \mathcal{D}[\ah](\check{\rho})
    \notag\\
    &+
    \kappa \ah[\check{\rho},\hat{\alpha}^\dagger]
    +
    \kappa [\hat{\alpha},\check{\rho}]\ad
    +
    \kappa \mathcal{D}[\hat{\alpha}](\check{\rho}),
    \label{eqn:L0t_appendix}
\end{align}
where
\begin{align}
 \omega^{D}_q
=
\delta_q
+
(\delta_a+\chi_g)|\alpha_g|^2
-
(\delta_a+\chi_h)|\alpha_h|^2 .
\end{align}
This zeroth-order dynamics therefore admits the steady-state manifold
\begin{align}
    \check{\rho}_{ss}^{(0)}
    =
    \left(
    \tilde P_g\ketbra{g}{g}
    +
    \tilde P_h\ketbra{h}{h}
    \right)\otimes\ketbra{0}{0},
\end{align}
which corresponds in the original frame to
\begin{align}
    \rhoh_{ss}^{(0)}
    =
    \tilde{P}_g \ketbra{g}{g}\otimes\ketbra{\alpha_g}{\alpha_g}
    +
    \tilde{P}_h \ketbra{h}{h}\otimes\ketbra{\alpha_h}{\alpha_h}.
\end{align}

In the weak-coupling regime $\abs{g_{\rm eff}}\abs{\alpha_g}^2, \abs{g_{\rm eff}}\abs{\alpha_h}^2 \ll \kappa$, the resonator relaxes on a faster timescale than the induced qubit transition. The dynamics therefore remains close to the zeroth-order steady-state manifold above, while the two-photon term $\mathcal{L}_1$ induces slow transitions between its $g$- and $h$-branch components. In the same displaced frame, the two-photon coupling becomes
\begin{align}
    \check{H}^{(1)}
    =
    g_{\rm eff}\bigl(\hat{A}^{(2)}\bigr)^\dagger \ketbra{g}{h}
    +
    g_{\rm eff}^* \hat{A}^{(2)} \ketbra{h}{g},
\end{align}
with
\begin{align}
    \hat{A}^{(2)}
    =
    \hat{D}^\dagger(\alpha_h)\ah^2 \hat{D}(\alpha_g).
\end{align}

Writing the displaced density matrix in the qubit basis as
\begin{align}
    \check{\rho}
    =
    \check{\rho}_{gg}\ketbra{g}{g}
    +
    \check{\rho}_{hh}\ketbra{h}{h}
    +
    \check{\rho}_{hg}\ketbra{h}{g}
    +
    \check{\rho}_{gh}\ketbra{g}{h},
\end{align}
the zeroth-order Liouvillian preserves the diagonal and off-diagonal sectors separately, while $\check{\mathcal L}_1$ couples them through $\check{\Hh}^{(1)}$. The population dynamics in the diagonal sector is therefore induced at second order through the intermediate coherences $\check{\rho}_{hg}$ and $\check{\rho}_{gh}$.

The first-order equation for the $hg$ coherence block reads
\begin{align}
    \dot{\check{\rho}}_{hg}^{(1)}
    =
    \check{\mathcal L}_{0,hg}\check{\rho}_{hg}^{(1)}
    -
    i g_{\rm eff}^*
    \left(
        \tilde P_g \hat{A}^{(2)}\ketbra{0}{0}
        -
        \tilde P_h \ketbra{0}{0}\hat{A}^{(2)}
    \right),
\end{align}
where $\check{\mathcal L}_{0,hg}$ is the zeroth-order Liouvillian acting in the $\ketbra{h}{g}$ coherence sector,
\begin{align}
    \check{\mathcal L}_{0,hg}(\cdot)
    \equiv
    \bra{h}\check{\mathcal L}_0\!\left[\ketbra{h}{g}\otimes(\cdot)\right]\ket{g}.
\end{align} Since this coherence relaxes faster than the population dynamics, we set $\dot{\check{\rho}}_{hg}^{(1)}\approx 0$, which gives
\begin{align}
    \check{\rho}_{hg}^{(1)}
    =
    i g_{\rm eff}^*
    \left(\check{\mathcal L}_{0,hg}^{-1}\right)
    \left(
        \tilde P_g \hat{A}^{(2)}\ketbra{0}{0}
        -
        \tilde P_h \ketbra{0}{0}\hat{A}^{(2)}
    \right).
\end{align}

The population dynamics then follows from the second-order equation
\begin{align}
    \dot{\check{\rho}}^{(2)}
    =
    \check{\mathcal L}_0\check{\rho}^{(2)}
    +
    \check{\mathcal L}_1\check{\rho}^{(1)}.
\end{align}
Projecting onto the slow subspace
$\{
\ketbra{g}{g}\otimes\ketbra{0}{0},
\ketbra{h}{h}\otimes\ketbra{0}{0}
\}$,
the term \(\check{\mathcal L}_1\check{\rho}^{(1)}\) induces population transfer through the intermediate coherence \(\check{\rho}_{hg}^{(1)}\). In particular,
\begin{align}
    \dot{\tilde P}_g
    &=
    \operatorname{Tr}\!\left[
    \left(\ketbra{g}{g}\otimes\ketbra{0}{0}\right)
    \check{\mathcal L}_1\check{\rho}^{(1)}
    \right]
    \notag\\
    &=
    2\Re\!\left\{
    -i g_{\rm eff}
    \operatorname{Tr}_r
    \left[
        \hat{A}^{(2)\dagger}
        \check{\rho}_{hg}^{(1)}
    \right]
    \right\}.
\end{align}
Here $\operatorname{Tr}_r$ denotes the partial trace over the resonator subspace.

Substituting the expression for $\check{\rho}_{hg}^{(1)}$ then yields the population rate equation quoted in Eq.~\eqref{eqn:Pevolution},
\begin{align}
    \dot{\tilde P}_g
    &=
    -\gamma_g \tilde P_g+\gamma_h \tilde P_h,
    \\
    \dot{\tilde P}_h
    &=
    -\dot{\tilde P}_g,
\end{align}
with
\begin{align}
    \gamma_g
    &=-
    2\abs{g_{\rm eff}}^2
    \Re \left\{
    \operatorname{Tr}_r
    \left[
    \hat{A}^{(2)\dagger}
   \check{\mathcal L}_{0,hg}^{-1} \left(
    \hat{A}^{(2)}\ketbra{0}{0}\right)
    \right]\right\},
    \notag\\
    \gamma_h
    &=-
    2\abs{g_{\rm eff}}^2
    \Re
      \left\{\operatorname{Tr}_r
    \left[
   \hat{A}^{(2)\dagger}
  \check{\mathcal L}_{0,hg}^{-1}
    \left(\ketbra{0}{0}\hat{A}^{(2)}\right)
    \right]\right\}.
    \label{eqn:gamma_appendix}
\end{align}
The rates can be evaluated by expanding in the photon-number basis and solving the resulting recursive relations. Details are provided in the Supplemental Material~\cite{note1}.

Using $\tilde P_g+\tilde P_h=1$, Eq.~\eqref{eqn:Pevolution} reduces to Eq.~\eqref{eqn:Prelax}. The corresponding analytical solution is
\begin{align}
    \tilde P_g(t)
    =
    \tilde P_g^{\rm ss}
    +
    \left[
        \tilde P_g(0)-\tilde P_g^{\rm ss}
    \right]
    e^{-\gamma t},
\end{align}
with $\tilde P_h(t)=1-\tilde P_g(t)$. This rate equation faithfully reproduces the dynamics of the reduced quantum model beyond the initial transient, while small deviations can appear at earlier times due to the adiabatic-elimination approximation.

\section{Resolved and Unresolved Regimes of the Transition Rates}
\label{app:regimes}

In the main text, we introduced the near-resonant distinguishability $D_{\rm cross}^{(g)}$ to characterize the resonator-state separation relevant to the upward transition from the computational subspace. For completeness, the corresponding quantity for the downward transition is
\begin{align}
D_{\rm cross}^{(h)}
\equiv
D\Big|_{\bar n_h=n_{\rm cross}^{(h)}}
=
n_{\rm cross}^{(h)}
\frac{(\chi_g-\chi_h)^2}
{(\delta_a+\chi_g)^2+\kappa^2/4}.
\end{align}
These quantities distinguish two limiting forms of the transition profiles, which we discuss below.

\subsection{Gaussian-like Profile in the Resolved Regime}

We first consider the regime in which the qubit-conditioned resonator states remain distinguishable near both branch-dependent transition points,
\begin{align}
D_{\rm cross}^{(g)} \gg 1,
\qquad
D_{\rm cross}^{(h)} \gg 1.
\end{align}
In this case, the upward and downward transitions are distributed over branch-dependent photon-number channels and must be treated separately.

We approximate these qubit-branch-resolved transitions using Fermi's golden rule. In the steady state, each qubit branch is associated with a conditional coherent resonator state, leading to the photon-number distributions
\begin{align}
P_g(n;n_g)=e^{-n_g}\frac{n_g^n}{n!},
\qquad
P_h(n;n_h)=e^{-n_h}\frac{n_h^n}{n!},
\end{align}
with $n_g=|\alpha_g|^2$ and $n_h=|\alpha_h|^2$. The two-photon interaction then couples the photon-number channels
\begin{align}
\ket{g,n} \to \ket{h,n-2},
\qquad
\ket{h,n} \to \ket{g,n+2}.
\end{align}
For these processes, the squared matrix elements are
\begin{align}
\left|
\bra{h,n-2}
\hat H^{(1)}
\ket{g,n}
\right|^2
&=
|g_{\rm eff}|^2 n(n-1),
\\
\left|
\bra{g,n+2}
\hat H^{(1)}
\ket{h,n}
\right|^2
&=
|g_{\rm eff}|^2 (n+1)(n+2).
\end{align}

The corresponding branch-dependent detunings are
\begin{align}
\Delta_g(n)
&=
\delta_q-2(\delta_a+\chi_h)-(\chi_g-\chi_h)n,
\\
\Delta_h(n)
&=
\delta_q-2(\delta_a+\chi_g)-(\chi_g-\chi_h)n,
\end{align}
 and their zeros define the crossing photon numbers
\begin{align}
n_{\rm cross}^{(g)}
&=
\frac{\delta_q-2(\delta_a+\chi_h)}{\chi_g-\chi_h},
\\
n_{\rm cross}^{(h)}
&=
\frac{\delta_q-2(\delta_a+\chi_g)}{\chi_g-\chi_h}
=
n_{\rm cross}^{(g)}-2.
\end{align}

For large crossing photon numbers, the transition is concentrated near the resonant photon-number channels, while the Poisson distributions vary smoothly over neighboring channels around the corresponding transition points. In this regime, the channel index $n$ can be treated as a continuous variable, and the branch-resolved transition rates can be approximated by the continuous-limit form of Fermi’s golden rule,
\begin{align}
\gamma_g(n_g)
&\approx
2\pi |g_{\rm eff}|^2
\int dn\,
P_g(n;n_g)\,
n(n-1)\,
\delta\!\bigl(\Delta_g(n)\bigr),
\notag
\\
\gamma_h(n_h)
&\approx
2\pi |g_{\rm eff}|^2
\int dn\,
P_h(n;n_h)\,
(n+1)(n+2)\,
\delta\!\bigl(\Delta_h(n)\bigr).
\label{eq:Ch_cont}
\end{align}
Since
\begin{align}
\frac{d\Delta_g}{dn}
=
\frac{d\Delta_h}{dn}
=
-(\chi_g-\chi_h),
\end{align}
these integrals give
\begin{align}
\gamma_g(n_g)
&\approx
\frac{2\pi |g_{\rm eff}|^2}{|\chi_g-\chi_h|}
\,n_{\rm cross}^{(g)}\!\bigl(n_{\rm cross}^{(g)}-1\bigr)\,
P_g\!\bigl(n_{\rm cross}^{(g)};n_g\bigr),
\\
\gamma_h(n_h)
&\approx
\frac{2\pi |g_{\rm eff}|^2}{|\chi_g-\chi_h|}
\,\bigl(n_{\rm cross}^{(h)}+1\bigr)\bigl(n_{\rm cross}^{(h)}+2\bigr)\,
P_h\!\bigl(n_{\rm cross}^{(h)};n_h\bigr).
\end{align}

Finally, for large photon number, the Poisson distribution is approximated by
\begin{align}
P_i(n;n_i)
\approx
\frac{1}{\sqrt{2\pi n_i}}
\exp\!\left[
-\frac{(n-n_i)^2}{2n_i}
\right],
\qquad
i=g,h.
\end{align}
Substituting the large-photon-number form of the Poisson distribution then gives the approximate Gaussian-like expressions for the transition rates
\begin{align}
\gamma_g(n_g)
&\approx
\tfrac{2\pi |g_{\rm eff}|^2}{|\chi_g-\chi_h|}
\tfrac{n_{\rm cross}^{(g)}\!\bigl(n_{\rm cross}^{(g)}-1\bigr)}
{\sqrt{2\pi n_g}}
\exp\!\left[
-\tfrac{\bigl(n_g-n_{\rm cross}^{(g)}\bigr)^2}{2n_g}
\right],\notag
\\
\gamma_h(n_h)
&\approx
\tfrac{2\pi |g_{\rm eff}|^2}{|\chi_g-\chi_h|}
\tfrac{\bigl(n_{\rm cross}^{(h)}+1\bigr)\bigl(n_{\rm cross}^{(h)}+2\bigr)}
{\sqrt{2\pi n_h}}
\exp\!\left[
-\tfrac{\bigl(n_h-n_{\rm cross}^{(h)}\bigr)^2}{2n_h}
\right].
\end{align}

The transition rates therefore acquire Gaussian-like profiles peaked near $n_{\rm cross}^{(g)}$ and $n_{\rm cross}^{(h)}$, with widths of order $\sqrt{n_{\rm cross}^{(g)}}$ and $\sqrt{n_{\rm cross}^{(h)}}$. This provides a useful smooth approximation to the transition profiles in the resolved regime, reflecting the Poisson photon-number statistics of the qubit-conditioned coherent states together with the resonance selection of the branch-resolved channels. The exact transition-rate expression in Eq.~\eqref{eqn:gamma_appendix} retains the full quantitative structure beyond this approximation, and numerical comparisons between the exact rates and the Gaussian approximation are provided in the Supplemental Material~\cite{note1}.

\subsection{Lorentzian Profile in the Unresolved Regime}

We next consider the opposite regime, in which the two qubit-conditioned resonator states strongly overlap near the two-photon transition points,
\begin{align}
D_{\rm cross}^{(g)} \ll 1,
\qquad
D_{\rm cross}^{(h)} \ll 1.
\end{align}
In this limit, the branch-dependent resonator states are no longer well resolved during the transition, and the dynamics is instead governed by a common effective field. The upward and downward processes therefore lose their branch-resolved structure and approach a common transition profile.

To describe this limit, we introduce the common coherent-state amplitude
\begin{align}
\alpha_c
=
-\frac{\varepsilon_d}{\delta_a-i\kappa/2},
\qquad
\bar n_c=|\alpha_c|^2,
\end{align}
which approximates both branch-dependent amplitudes when the resonator-state separation is small,
\begin{align}
\alpha_g \approx \alpha_h \approx \alpha_c.
\end{align}
The two-photon interaction then acts on a single coherent field, giving the effective matrix element
\begin{align}
\left|
\bra{h,\alpha_c}
\hat H^{(1)}
\ket{g,\alpha_c}
\right|^2
=
|g_{\rm eff}|^2 \bar n_c^2.
\end{align}

The residual separation between the two qubit-conditioned resonator states produces measurement-induced dephasing, characterized by
\begin{align}
\Gamma_\phi
=
\frac{\kappa}{2}|\alpha_g-\alpha_h|^2.
\end{align}
The corresponding detuning is determined by the common resonator photon number,
\begin{align}
\Delta_c
=
\delta_q-(\chi_g-\chi_h)\bar n_c.
\end{align}
The transition rates therefore acquire a Lorentzian profile,
\begin{align}
\gamma
&\approx
2 \pi |g_{\rm eff}|^2 \bar n_c^2
\frac{1}{\pi}
\frac{\Gamma_\phi}{\Delta_c^2+\Gamma_\phi^2}
\notag\\
&=
2 |g_{\rm eff}|^2 \bar n_c^2
\frac{\Gamma_\phi}
{\bigl[\delta_q-(\chi_g-\chi_h)\bar n_c\bigr]^2+\Gamma_\phi^2}.
\end{align}
This profile is centered at the common unresolved crossing photon number $n_{\rm cross}^{\rm (unres)}\equiv\delta_q/(\chi_g-\chi_h)$ and gives a pronounced rate peak near resonance. In contrast to the resolved regime, where the transition is distributed over many branch-resolved photon-number channels, this behavior reflects the strong overlap of the qubit-conditioned resonator states and approaches a more semiclassical-like sharp transition profile. The same Lorentzian form can also be obtained by expanding the exact transition-rate expressions in the small-distinguishability limit, as shown in the Supplemental Material~\cite{note1}.

\section{Extension to Higher-Order Multi-Photon Resonances}
\label{app:multiphonon_extension}

The reduced driven-dissipative framework developed in this work extends naturally to a generic $k$-photon resonance between two relevant qubit levels, denoted by $\ket{\mu}$ and $\ket{\nu}$, satisfying $\omega_\nu-\omega_\mu \approx k\omega_r$. Following the same SW construction and rotating-wave approximation as in Appendix~\ref{app:reduced_framework} and retaining only the minimal near-resonant structure relevant to the $k$-photon transition, the effective Hamiltonian takes the form
\begin{align}
\hat H_{\rm eff}^{(k)}
=&\,
\delta_a \hat a^\dagger \hat a
+
(\delta_\mu+\chi_\mu \hat a^\dagger \hat a)\ketbra{\mu}{\mu}
+
(\delta_\nu+\chi_\nu \hat a^\dagger \hat a)\ketbra{\nu}{\nu}
\notag\\
&+
g_{\rm eff}^{(k)}(\hat a^\dagger)^k \ketbra{\mu}{\nu}
+
\bigl(g_{\rm eff}^{(k)}\bigr)^* \hat a^k \ketbra{\nu}{\mu}
+
\varepsilon_d(\hat a+\hat a^\dagger),
\label{eq:Heff_kph_appendix}
\end{align}
where $\delta_{\mu/\nu}$ are the shifted energies, $\chi_{\mu/\nu}$ are the dispersive couplings, and $g_{\rm eff}^{(k)}$ is the effective $k$-photon coupling strength.

Including resonator loss, the open-system dynamics is again governed by a Lindblad master equation of the same form as Eq.~\eqref{eqn:MasterEffective}. In the weak-coupling limit $|g_{\rm eff}^{(k)}|\ll\kappa$, the dynamics can be analyzed using the same adiabatic-elimination procedure as in Appendix~\ref{app:rate_equations}.

At zeroth order, the steady-state manifold is spanned by qubit-state-conditioned coherent states,
\begin{align}
\hat\rho_{ss}^{(0)}
=
\tilde P_\mu \ketbra{\mu}{\mu}\otimes\ketbra{\alpha_\mu}{\alpha_\mu}
+
\tilde P_\nu \ketbra{\nu}{\nu}\otimes\ketbra{\alpha_\nu}{\alpha_\nu},
\end{align}
where the conditional coherent-state amplitudes are
\begin{align}
\alpha_{\mu/\nu}
=
-\frac{\varepsilon_d}{\delta_a+\chi_{\mu/\nu}-i\kappa/2}.
\end{align}

Transforming to the corresponding conditional displaced frame, the relevant transition operator becomes
\begin{align}
\hat A^{(k)}
=
\hat D^\dagger(\alpha_\nu)\hat a^k \hat D(\alpha_\mu),
\end{align}
which generalizes the two-photon operator $\hat{A}^{(2)}$ introduced in Appendix~\ref{app:rate_equations}. Treating the dynamics perturbatively in the weak-coupling limit yields a population rate equation of the same form as Eq.~\eqref{eqn:Pevolution},
\begin{align}
\dot{\tilde P}_\mu
=
-\gamma_\mu^{(k)} \tilde P_\mu
+
\gamma_\nu^{(k)} \tilde P_\nu,
\qquad
\dot{\tilde P}_\nu
=
-\dot{\tilde P}_\mu.
\end{align}

The corresponding transition rates take the form
\begin{align}
\gamma_\mu^{(k)}
&=
-2\abs{g_{\rm eff}^{(k)}}^2
\Re\!\left\{
\operatorname{Tr}_r
\left[
\hat A^{(k)\dagger}
\check{\mathcal L}_{0,\nu\mu}^{-1}
\bigl(
\hat A^{(k)}\ketbra{0}{0}
\bigr)
\right]
\right\},
\notag\\
\gamma_\nu^{(k)}
&=
-2\abs{g_{\rm eff}^{(k)}}^2
\Re\!\left\{
\operatorname{Tr}_r
\left[
\hat A^{(k)\dagger}
\check{\mathcal L}_{0,\nu\mu}^{-1}
\bigl(
\ketbra{0}{0}\hat A^{(k)}
\bigr)
\right]
\right\},
\end{align}
where $\check{\mathcal L}_{0,\nu\mu}$ denotes the zeroth-order Liouvillian in the $\ketbra{\nu}{\mu}$ coherence sector. As in the two-photon case, these rates can be evaluated by expanding in the photon-number basis and solving the resulting recursive relations. Further derivation details for the higher-order effective Hamiltonian and the corresponding transition-rate expressions are provided in the Supplemental Material~\cite{note1}.

The $k$-photon resonance condition is determined by the qubit-state-dependent crossing photon numbers. For the upward transition
\begin{align}
\ket{\mu,n}\rightarrow \ket{\nu,n-k},
\end{align}
the resonance condition gives
\begin{align}
n_{\rm cross}^{(\mu)}
=
\frac{\delta_{\nu\mu}-k(\delta_a+\chi_\nu)}{\chi_\mu-\chi_\nu},
\qquad
\delta_{\nu\mu}\equiv\delta_\nu-\delta_\mu.
\end{align}
For the downward transition
\begin{align}
\ket{\nu,n}\rightarrow \ket{\mu,n+k},
\end{align}
the corresponding crossing photon number is
\begin{align}
n_{\rm cross}^{(\nu)}
=
\frac{\delta_{\nu\mu}-k(\delta_a+\chi_\mu)}{\chi_\mu-\chi_\nu}
=
n_{\rm cross}^{(\mu)}-k.
\end{align}
This driven-dissipative description and rate-equation structure extend naturally to higher-order measurement-induced transitions between relevant near-resonant levels and apply broadly to different superconducting qubit architectures.

\footnotetext{See Supplemental Material for details\label{note1}}

\bibliographystyle{apsrev4-2}
\bibliography{MIST}

\end{document}


\title{Supplemental Material:\\{Driven-Dissipative Dynamics of Measurement-Induced State Transitions}
}
\author{Bo-Syun Pan}
\affiliation{Department of Physics and Center for Theoretical Physics, National Taiwan University, Taipei 106319, Taiwan}

\author{Yen-Hsiang Lin}
\affiliation{Department of Physics, National Tsing Hua University, Hsinchu 300013, Taiwan}
\affiliation{National Center for Excellence in Quantum Information Science and Engineering, National Tsing Hua University, Hsinchu 300013, Taiwan}

\author{Chiao-Hsuan Wang}
\email{chiaowang@phys.ntu.edu.tw}
\affiliation{Department of Physics and Center for Theoretical Physics, National Taiwan University, Taipei 106319, Taiwan}
\affiliation{Center for Quantum Science and Engineering, National Taiwan University, Taipei 106319, Taiwan}
\affiliation{Physics Division, National Center for Theoretical Sciences, Taipei 106319, Taiwan}

\maketitle
\section{Detailed Derivations of the Reduced Quantum Model}
This section provides the detailed algebra underlying Appendix A of the main text, including the explicit Schrieffer--Wolff recursion, the transformation of the drive and dissipator, and the extension to higher-order multiphoton resonances.
We derive the reduced qubit–resonator Hamiltonian under the two-photon resonance condition, starting from the Hamiltonian of the full quantum model of a superconducting qubit coupled to a classically driven readout resonator,  
\begin{align}
    \Hh_{\rm{full}}(t) = \Hh_{qr} + \Hh_{d}(t).
\end{align}
Note that we take $\hbar = 1$ throughout for simplicity.

The static qubit–resonator Hamiltonian is
\begin{align}
    \hat{H}_{qr} = \omega_r \ad \ah + \hat{H}_q + ig \hat{n} (\ad - \ah),
    \label{eqn:FullQubitResonatorHamiltonian}
\end{align}
where $\omega_r$ is the bare resonator frequency, $g$ is the qubit–resonator coupling strength, and $\hat{n}$ is the qubit charge operator.

The qubit Hamiltonian reads
\begin{align}
    \hat{H}_q = 4E_C \hat{n}^2 + \frac{1}{2} E_L \hat{\varphi}^2 - E_J \cos(\hat{\varphi} - \varphi_{\rm{ext}}),
\end{align}
where $E_C$ is the charging energy, $E_L$ is the inductive energy, $E_J$ is the Josephson energy, $\hat{\varphi}$ is the superconducting phase operator, and $\varphi_{\rm{ext}}$ is the externally applied flux in units of the flux quantum $\Phi_0$.  
This Hamiltonian can be diagonalized as
\begin{align}
    \hat{H}_q = \sum_{j} \omega_{j} \ketbra{j}{j},
\end{align}
where $\omega_{j}$ is the eigenfrequency of the $j$-th qubit eigenstate $\ket{j}$.

The resonator is driven by a classical field,
\begin{align}
    \hat{H}_d(t) = -2 i \varepsilon_d(t) \sin(\omega_d t) (\ad - \ah),
\end{align}
where $\varepsilon_d(t)$ and $\omega_d$ are the drive amplitude and frequency, respectively.  
All operators and parameters follow the definitions in the main text.

Including photon loss in the resonator, the system dynamics is governed by the master equation
\begin{equation}
    \dot{\rhoh} = -i[\hat{H}_{\rm full}, \rhoh] + \kappa\,\mathcal{D}[\ah](\rhoh),
    \label{eqn:FullMasterEquation}
\end{equation}
with the Lindblad dissipator defined as 
$\mathcal{D}[\hat{o}](\rho) = \hat{o}\rho\hat{o}^\dagger - \tfrac{1}{2}\{\hat{o}^\dagger\hat{o},\,\rho\}$, 
where $\kappa$ denotes the resonator decay rate.

We assume that no qubit transition is resonant with the resonator at the single-photon level, i.e., $\abs{g n_{ij}}\ll \abs{\omega_j - \omega_i - \omega_r}$ for all $i,j$, where $n_{ij} \equiv \bra{i}\hat{n}\ket{j}$. In contrast, the qubit ground state $\ket{g}$ and an excited state $\ket{h}$ satisfy the two-photon resonance condition $\omega_h - \omega_g \approx 2\omega_r$, leading to two-photon processes mediated by virtual intermediate states. To describe the dynamics within the resonant subspace, we perform a standard Schrieffer–Wolff transformation~\cite{Schrieffer1966,Blais2007,Blais2021} to diagonalize the qubit–resonator Hamiltonian $\hat{H}_{qr}$ to $\mathcal{O}(g)$, which captures the effective two-photon interactions generated by virtual single-photon transitions.

\subsection{General Framework of the Schrieffer–Wolff Transformation}
\label{sec:Schrieffer--Wolff}
We set up a general Schrieffer–Wolff (SW) framework for diagonalizing the qubit–resonator Hamiltonian $\hat{H}_{qr}$ order by order in the coupling strength $g$. We first decompose the Hamiltonian as
\begin{align}
\hat{H}_{qr} = \hat{H}_0 + \hat{V}^{(1)},
\end{align}
where $\hat{H}_0 = \hat{H}_q + \omega_r\ad\ah$ is the free Hamiltonian of the uncoupled qubit and resonator, whose eigenstates define the bare-state basis, and
$\hat{V}^{(1)} = ig \hat{n} (\ad - \ah)$ is the first-order interaction term.

The SW transformation is defined as
\begin{align}
    &\hat{H}_{qr}\rightarrow e^{\hat{S}}\hat{H}_{qr} e^{-\hat{S}},
\end{align}
where the anti-Hermitian generator is expanded up to order $\mathcal{O}(g^N)$ as $\hat{S} = \sum_{p = 1}^{N}\Sh{p}$, with each $\Sh{p}$ being of order $\mathcal{O}(g^p)$.
The generator is chosen to cancel all off-diagonal terms through $\mathcal{O}(g^N)$, so that the transformed Hamiltonian takes the form
\begin{align}
    e^{\hat{S}}\hat{H}_{qr}e^{-\hat{S}} = \hat{H}_0 + \left(\sum_{p=1}^{N}\Vh{p}\right)_{\rm d.} + \Vh{N+1} + \mathcal{O}(g^{N+2}),
    \label{eqn:HNdiagonal}
\end{align}
where $\Vh{p}$ denotes the $\mathcal{O}(g^p)$ effective interaction obtained after eliminating the off-diagonal terms up to order $\mathcal{O}(g^{p-1})$.
For any operator $\hat{O}$, we denote its diagonal and off-diagonal parts by $(\hat{O})_{\rm d.}$ and $(\hat{O})_{\rm o.d.}$, respectively.

To obtain a recursive construction for $\hat{S}^{(p)}$ and $\hat{V}^{(p)}$, we use the Zassenhaus formula \cite{Magnus1954}
\begin{align}
    e^{t(\hat{X}+\hat{Y})} = e^{t\hat{X}}e^{t\hat{Y}}e^{-\frac{t^2}{2}[\hat{X}, \hat{Y}]} \cdots,
\end{align}
which leads to
\begin{align}
    &\exp\left(\hat{S}^{(N)} +\sum_{p=1}^{N-1}\hat{S}^{(p)} \right) = \exp\left(\hat{S}^{(N)}\right)\exp\left(\sum_{p=1}^{N-1}\hat{S}^{(p)}\right)\left(1+\mathcal{O}(g^{N+1})\right),\\
    &e^{\hat{S}}\hat{H}_{qr}e^{-\hat{S}} 
   = \hat{H}_0 + \sum_{p = 1}^{N-1}(\hat{V}^{(p)})_{\rm d.} + \hat{V}^{(N)} + [\hat{S}^{(N)}, \hat{H}_{0}] + \mathcal{O}(g^{N+1}).
    \label{eqn:SWGeneralHamiltonianExpansion}
\end{align}
To match the diagonalized form up to $\mathcal{O}(g^N)$, this requires that for all $N \geq 1$,
\begin{align}
    [\hat{S}^{(N)}, \hat{H}_0] = -\left(\hat{V}^{(N)}\right)_{\rm o.d.} .
    \label{eqn:SWGeneratorCondition}
\end{align}
Starting from the explicit forms of $\hat{H}_0$ and $\hat{V}^{(1)}$ defined above, 
the recursion relation in Eq.~\eqref{eqn:SWGeneratorCondition} 
allows $\hat{S}^{(N)}$ and $\hat{V}^{(N+1)}$ to be constructed iteratively for all $N \geq 1$.

For an explicit construction, we use the normal ordered expansions
\begin{align}
    \hat{S}^{(p)} = \sum_{nm} \hat{S}^{(p)}_{nm}\left(\ad\right)^{n}\ah^{m}, \quad \hat{V}^{(p)} = \sum_{nm} \hat{V}^{(p)}_{nm}\left(\ad\right)^n\ah^m,
    \label{eqn:NormalOrdering}
\end{align}
where $\ShNO{p}{n}{m}$ and $\VhNO{p}{n}{m}$ are operator components acting in the qubit subspace. With $\hat{S}^{(p)}$ anti-Hermitian and $\hat{V}^{(p)}$ Hermitian,
\begin{align}
    \left(\ShNO{p}{n}{m}\right)^{\dagger} = -\ShNO{p}{m}{n}, \quad \left(\VhNO{p}{n}{m}\right)^{\dagger} = \VhNO{p}{m}{n}.
    \label{eqn:NormalOrderingConstraints}
\end{align}
Combining Eqs.~\eqref{eqn:NormalOrdering} and \eqref{eqn:SWGeneratorCondition}, the coefficients of $\hat S^{(p)}$ for a given order $p$ are
\begin{align}
    \bra{i}\ShNO{p}{n}{m}\ket{j} = -\frac{\bra{i}(\VhNO{p}{n}{m})_{\rm o.d.}\ket{j}}{\left(\omega_{ij} - \omega_r(n-m) \right)} = -\frac{\bra{i}\VhNO{p}{n}{m}\ket{j}}{\left(\omega_{ij} - \omega_r(n-m) \right)}\left(1 - \delta_{ij}\delta_{nm}\right),
    \label{eqn:SWGeneratorSolution}
\end{align}
where $\delta_{ij}$ is the Kronecker delta symbol, and $\omega_{ij} = \omega_j - \omega_i$ denotes the frequency difference between qubit eigenstates $\ket{i}$ and $\ket{j}$. Note that $\hat V^{(p)}$ is provided by the $(p{-}1)$-th order results. This offers a recursive procedure for constructing $\hat{S}$ and the effective interactions order by order.

\subsection{Two-Photon Resonance Case}
\label{sec:SWTwoPhoton}
To obtain the effective two-photon interaction, we perform the first-order SW transformation, which generates the effective Hamiltonian up to second order in $g$. From the first-order interaction $\hat{V}^{(1)} = ig\hat{n}(\ad - \ah)$, the corresponding first-order generator reads

\begin{align}
    \Sh{1} = \ShNO{1}{1}{0}\ad + \ShNO{1}{0}{1}\ah, \quad \bra{i}\ShNO{1}{1}{0}\ket{j} = -\frac{ign_{ij}}{(\omega_{ij} - \omega_r)}, \quad \ShNO{1}{0}{1} = -\left(\ShNO{1}{1}{0}\right)^{\dagger}.
    \label{eqn:FirstOrderGenerator}
\end{align}
The second-order effective interaction $\hat{V}^{(2)}$ then follows as
\begin{align}
    \hat{V}^{(2)} &= \frac{1}{2}[\hat{S}^{(1)}, [\hat{S^{(1)}}, \hat{H}_0]] + [\hat{S}^{(1)}, \hat{V}^{(1)}] = \frac{1}{2}[\hat{S}^{(1)}, \hat{V}^{(1)}] \notag \\
    & = \VhNO{2}{1}{1}\ad\ah + \VhNO{2}{0}{0} + \VhNO{2}{2}{0}(\ad)^2 + \VhNO{2}{0}{2}\ah^2,
    \label{eqn:SecondOrderInteraction}
\end{align}
with the explicit forms
\begin{align}
    \VhNO{2}{1}{1} &= \frac{1}{2}\left([\ShNO{1}{1}{0}, \VhNO{1}{0}{1}] + [\ShNO{1}{0}{1}, \VhNO{1}{1}{0}]\right) = g^2\sum_{i, j, k=0}^\infty \ketbra{i}{j}n_{ik}n_{kj}\left(\frac{\omega_{ki}}{\omega_{ki}^2 - \omega_r^2} + \frac{\omega_{kj}}{\omega_{kj}^2-\omega_r^2}\right), \\
    \VhNO{2}{0}{0} &= \frac{1}{2}\left(\ShNO{1}{0}{1}\VhNO{1}{1}{0} - \VhNO{1}{0}{1}\ShNO{1}{1}{0}\right) = \frac{g^2}{2}\sum_{i, j, k=0}^{\infty}\ketbra{i}{j} n_{ik}n_{kj}\left(\frac{1}{\omega_{ki} - \omega_r} + \frac{1}{\omega_{kj} - \omega_r}\right),\\
    \VhNO{2}{2}{0} &= \frac{1}{2}[\ShNO{1}{1}{0}, \VhNO{1}{1}{0}] = \frac{g^2}{2}\sum_{i, j, k=0}^{\infty}\ketbra{i}{j} n_{ik}n_{kj}\left(\frac{1}{\omega_{ik} - \omega_r} - \frac{1}{\omega_{kj} - \omega_r}\right),\\
    \VhNO{2}{0}{2} &= \left(\VhNO{2}{2}{0}\right)^\dagger.
\end{align}
The transformed Hamiltonian takes the form
\begin{align}
    e^{\hat{S}}\hat{H}_{qr}e^{-\hat{S}} = \hat{H}_0 + \Vh{2} + \mathcal{O}(g^3) = \hat{H}_q + \omega_r\ad\ah + \VhNO{2}{1}{1}\ad\ah + \VhNO{2}{0}{0} + \VhNO{2}{2}{0}(\ad)^2 + \left(\VhNO{2}{2}{0}\right)^\dagger \ah^2 + \mathcal{O}(g^3).
\end{align}

We now restrict the qubit Hilbert space to the subspace spanned by $\ket{g}, \ket{h}$ and apply the rotating wave approximation (RWA) to neglect non-resonant terms. For $\VhNO{2}{1}{1}$ and $\VhNO{2}{0}{0}$, we retain only their diagonal components, giving state-dependent dispersive and energy shifts. For $\VhNO{2}{2}{0}$, we keep only the terms involving $\ketbra{g}{h}$, which describe the resonant two-photon transition. The transformed Hamiltonian under RWA is then
\begin{align}
    e^{\hat{S}}\hat{H}_{qr}e^{-\hat{S}} &\approx \omega_r\ad\ah + (\omega_g + \Lambda_g + \chi_g\ad\ah)\ketbra{g}{g} + (\omega_h + \Lambda_h + \chi_h\ad\ah)\ketbra{h}{h} \notag\\ 
    &\quad + g_{\rm eff}(\ad)^2\ketbra{g}{h} + {g^*_{\rm eff}}(\ah)^2\ketbra{h}{g},
\end{align}
where $\chi_{i} = \bra{i}\VhNO{2}{1}{1}\ket{i}$ are the dispersive coupling strengths, $\Lambda_{i} = \bra{i}\VhNO{2}{0}{0}\ket{i}$ are the qubit energy shifts, and $g_{\rm eff} = \bra{g}\VhNO{2}{2}{0}\ket{h}$ is the effective two-photon coupling strength.

We next apply the same SW transformation to the drive Hamiltonian under a constant amplitude $\varepsilon_d(t)=\varepsilon_d$,
\begin{align}
e^{\hat{S}}\hat{H}_{d} e^{-\hat{S}} &= -2i\varepsilon_d\sin(\omega_d t)(\ad - \ah) + [\hat{S}, -2i\varepsilon_d\sin(\omega_d t)(\ad - \ah)] + \mathcal{O}(\varepsilon_d g^2) \notag\\
        &= -2i\varepsilon_d\sin(\omega_d t)(\ad - \ah) - 2i\varepsilon_d\sin(\omega_d t)(\ShNO{1}{0}{1}+ \ShNO{1}{1}{0}) + \mathcal{O}\left(\frac{\varepsilon_d g^2|n_{ij} n_{jk}|}{|(\omega_{ij} - \omega_r)(\omega_{jk} - \omega_r)|}\right).
\label{eqn:HdSW}
\end{align}
The third term can be neglected assuming $|g n_{ij}|\ll |\omega_{ij} - \omega_r|$.
The second term evaluates to
\begin{align}
    - 2i\varepsilon_d\sin(\omega_d t)(\ShNO{1}{0}{1}+ \ShNO{1}{1}{0}) = i\varepsilon_d(e^{i\omega_d t} - e^{-i\omega_d t})\left(\sum_{i, j} \frac{gn_{i j}}{\omega_{ij} - \omega_r}\ketbra{i}{j} + \text{H.C.}\right).
    \label{eqn:Hdsecond}
\end{align}
For a near-resonant drive $\omega_d \approx \omega_r$, with $|\omega_{ij}-\omega_d|$ remaining large for all qubit states, the terms in Eq.~\eqref{eqn:Hdsecond} are off-resonant and can be neglected provided
\begin{align}
    |\varepsilon_d g n_{i j}| \ll |\left(\omega_{ij} - \omega_r\right)\left(\omega_{ij} \pm \omega_d\right)|.
\end{align}
Under the RWA, the second term in Eq.~\eqref{eqn:HdSW} is dropped.

Hence, the transformed drive Hamiltonian can be simplified to:
\begin{align}
    e^{\hat{S}}\hat{H}_{d}e^{-\hat{S}} 
    \approx \varepsilon_d \ad e^{-i\omega_d t} + \varepsilon_d \ah e^{i\omega_d t}.
    \label{eqn:TransformedDriveHamiltonian}
\end{align}

Finally, we apply the SW transformation to the Lindblad dissipator,
\begin{align}
    \kappa\mathcal{D}[\ah](\rhoh) \rightarrow \kappa\mathcal{D}[e^{\hat{S}}\ah e^{-\hat{S}}](\rhoh) = \kappa \mathcal{D}[\ah](\rhoh) - \frac{\kappa}{2}\left\{\left(2\ShNO{1}{1}{0}\rhoh\ad - \ad \ShNO{1}{1}{0}\rhoh - \rhoh\ad\ShNO{1}{1}{0}\right) + \text{H.C.} \right\} + \mathcal{O}\left(\frac{\kappa g^2|n_{ij} n_{jk}|}{|(\omega_{ij} - \omega_r)(\omega_{jk} - \omega_r)|}\right).
\end{align}
Note that the density operator after the transformation is in the dressed-state basis. 

Same as in Eq.~\ref{eqn:HdSW}, the second order correction can be neglected assuming $|g n_{ij}|\ll |\omega_{ij} - \omega_r|$.
To estimate the first-order correction, we move to the rotating frame 
$\hat{U}(t)=\exp[-i(\hat{H}_q+\omega_d\ad\ah)t]$, $\omega_d \approx \omega_r $ under which the operator 
$\ShNO{1}{1}{0}\ad$ acquires fast oscillating phases 
$e^{-i(\omega_j-\omega_i-\omega_d)t}$. Since all detunings 
$|\omega_{ij}-\omega_r|$ are assumed large, the correction is suppressed and 
can be neglected when 
\begin{align}
    |\kappa g n_{ij}| \ll (\omega_{ij}-\omega_r)^2.
\end{align}

Collecting all terms under the RWA, the Hamiltonian in the reduced subspace, $\hat{H}_{\rm red} = e^{\hat{S}}\hat{H}_{\rm full}e^{-\hat{S}}$, and the corresponding Lindblad master equation read
\begin{align}
    \hat{H}_{\rm red} 
         & \approx \omega_r\ad\ah + (\omega_g + \Lambda_g + \chi_g\ad\ah)\ketbra{g}{g} + (\omega_h + \Lambda_h + \chi_h\ad\ah)\ketbra{h}{h}\notag\\
&+ g_{\rm eff}(\ad)^2\ketbra{g}{h} + {g^*_{\rm eff}}(\ah)^2\ketbra{h}{g} + \varepsilon_d\left(\ad e^{-i\omega_d t} + \ah e^{i\omega_d t}\right),\\
    \dot{\rhoh} & = -i[\hat{H}_{\rm red}, \rhoh] + \kappa\mathcal{D}[\ah](\rhoh).
    \label{eqn:ReducedMasterEquation}
\end{align}

Moving to the rotating frame $\hat{U}_r(t) = \exp[-i(\omega_d\ad\ah + 2\omega_d\ketbra{h}{h})t]$, we arrive at the effective Hamiltonian and the corresponding master equation presented in the main text,
\begin{align}
    \hat{H}_{\rm eff} &= \delta_a\ad\ah + (\delta_g + \chi_g\ad\ah)\ketbra{g}{g} + (\delta_h+\chi_h\ad\ah)\ketbra{h}{h}
+ g_{\rm eff}(\ad)^2\ketbra{g}{h} + {g^*_{\rm eff}}(\ah)^2\ketbra{h}{g} + \varepsilon_d\left(\ad + \ah\right),
        \label{eqn:Heff}\\
    \dot{\rhoh} & = -i[\hat{H}_{\rm eff}, \rhoh] + \kappa\mathcal{D}[\ah](\rhoh),
    \label{eqn:effMasterEquation}
\end{align}
where $\delta_a = \omega_r - \omega_d$, $\delta_g = \omega_g + \Lambda_g$, $\delta_h = \omega_h + \Lambda_h - 2\omega_d$.

\subsection{Extension to Higher-Order Resonances: Three-Photon Example}
As an illustration of higher-order processes, we present the extension of the SW transformation to the third order. The effective interaction at $\mathcal{O}(g^3)$ is
\begin{align}
    \Vh{3} & = [\Sh{2}, \Vh{1}] + \frac{1}{2!}\left([\Sh{2}, [\Sh{1}, \hat{H}_0]] + [\Sh{1}, [\Sh{1}, \Vh{1}]] 
    + [\Sh{1}, [\Sh{2}, \hat{H}_0]]\right)\notag\\
    & \quad + \frac{1}{3!}\left([\Sh{1}, [\Sh{1}, [\Sh{1}, \hat{H}_0]]]\right)\notag\\
    &=\frac{1}{2}[\Sh{2}, \Vh{1}] + \frac{1}{6}[\Sh{1}, \Vh{2}] + \frac{1}{2}[\Sh{1}, (\Vh{2})_{\rm d.}].
    \label{eqn:ThirdOrderInteraction}
\end{align}

By substituting the solution of the generator $\hat{S}^{(p)}$ from Eq.~\eqref{eqn:SWGeneratorSolution} into the expression for $\hat{V}^{(3)}$ in Eq.~\eqref{eqn:ThirdOrderInteraction}, the non-zero elements of $\hat{V}^{(3)}$ can be expressed in terms of lower-order contributions of $\hat{V}$ and $\hat{S}$:
\begin{align}
    \VhNO{3}{1}{0} &= \frac{1}{2}\left(\ShNO{2}{1}{1}\VhNO{1}{1}{0} + [\ShNO{2}{0}{0}, \VhNO{1}{1}{0}] - 2\VhNO{1}{0}{1}\ShNO{2}{2}{0}\right) + \frac{1}{6}\left(\VhNO{2}{1}{1}\ShNO{1}{1}{0}+[\ShNO{1}{1}{0}, \VhNO{2}{0}{0}] + 2\ShNO{1}{0}{1}\VhNO{2}{2}{0}\right)\notag\\ 
    &+ \frac{1}{2}\left((\VhNO{2}{1}{1})_{\rm d.}\ShNO{1}{1}{0} + [\ShNO{1}{1}{0}, (\VhNO{2}{0}{0})_{\rm d.}]\right),\\
    \VhNO{3}{2}{1} &= \frac{1}{2}\left([\ShNO{2}{1}{1}, \VhNO{1}{1}{0}] + [\ShNO{2}{2}{0}, \VhNO{1}{0}{1}]\right) + \frac{1}{6}\left([\ShNO{1}{1}{0}, \VhNO{2}{1}{1}] + [\ShNO{1}{0}{1}, \VhNO{2}{0}{2}]\right) + \frac{1}{2}\left([\ShNO{1}{1}{0}, (\VhNO{2}{1}{1})_{\rm d.}]\right),\\
    \VhNO{3}{3}{0} &= \frac{1}{2}[\ShNO{2}{2}{0}, \VhNO{1}{1}{0}] + \frac{1}{6}[\ShNO{1}{1}{0}, \VhNO{2}{0}{2}],\\
    \VhNO{3}{0}{1} &= \left(\VhNO{3}{1}{0}\right)^\dagger, \quad \VhNO{3}{1}{2} = \left(\VhNO{3}{2}{1}\right)^\dagger, \quad \VhNO{3}{0}{3} = \left(\VhNO{3}{3}{0}\right)^\dagger.
\end{align}

For a three-photon resonance between two relevant near-resonant levels $\ket{\mu}$ and $\ket{\nu}$, with no lower-order resonances, the reduced Hamiltonian after the rotating-wave approximation is
\begin{align}
    e^{\hat{S}}\hat{H}_{qr}e^{-\hat{S}}
    &\approx
    \hat{H}_q + \omega_r \ad\ah
    + (\omega_\mu + \Lambda_\mu + \chi_{\mu}\ad\ah)\ketbra{\mu}{\mu}
    + (\omega_\nu + \Lambda_\nu + \chi_{\nu}\ad\ah)\ketbra{\nu}{\nu}
    \notag\\
    &\quad
    + g_{\rm eff}^{(3)}(\ad)^{3}\ketbra{\mu}{\nu}
    + \bigl(g_{\rm eff}^{(3)}\bigr)^{*}(\ah)^{3}\ketbra{\nu}{\mu},
\end{align}
where
\begin{align}
g^{(3)}_{\rm eff} = \bra{\mu}\VhNO{3}{3}{0}\ket{\nu}.
\end{align}
The matrix elements of $\VhNO{3}{3}{0}$ are
\begin{align}
\bra{i}\VhNO{3}{3}{0}\ket{j} 
&= \sum_{k\ell} ig^3 n_{i\ell} n_{\ell k} n_{kj}
\left\{
\begin{aligned}
    &-\left( \frac{1}{2} \frac{1}{\omega_{ki} - 2\omega_r} - \frac{1}{6} \frac{1}{\omega_{kj} - \omega_r} \right)
     \left( \frac{1}{\omega_{i\ell} - \omega_r} - \frac{1}{\omega_{kj} - \omega_r} \right) \\[0.3em]
    &+ \left( \frac{1}{2} \frac{1}{\omega_{\ell j} - 2\omega_r} - \frac{1}{6} \frac{1}{\omega_{i\ell} - \omega_r} \right)
     \left( \frac{1}{\omega_{\ell k} - \omega_r} - \frac{1}{\omega_{kj} - \omega_r} \right)
\end{aligned}
\right\}.
\end{align}
The same construction generalizes to higher-order $k$-photon resonances, yielding effective couplings $g_{\rm eff}^{(k)}$ through the Schrieffer--Wolff transformation followed by the rotating-wave approximation.

\section{Driven-Dissipative Analysis of the Reduced Quantum Model}
\label{sec:steady_state_relaxation}
We consider the Hamiltonian in the reduced subspace, where the qubit states are spanned by ${\ket{g}, \ket{h}}$. The corresponding Pauli operators are defined as
\begin{align}
    \shz = \ketbra{h}{h} - \ketbra{g}{g}, \quad \shp = \ketbra{h}{g}, \quad \shm = \ketbra{g}{h}.
\end{align}
Treating $g_{\rm eff}$ as a small parameter, the Hamiltonian in Eq.~\eqref{eqn:Heff} can be expanded to the zeroth and first order in $g_{\rm eff}$ as
\begin{align}
    \hat{H}_{\rm eff} = \hat{H}^{(0)} + \hat{H}^{(1)}, \label{eq:HeffSimplify}
\end{align}
where
\begin{align}
    \hat{H}^{(0)} &= \check{\chi} \ad\ah +\frac{1}{2}\delta_q\shz +\varepsilon_d(\ad + \ah) + \frac{1}{2}(\delta_h + \delta_g) \label{eq:HeffZeroOrder}\\
    \hat{H}^{(1)} &= g_{\rm eff}{(\ad)}^2\shm + g_{\rm eff}^* \ah^2 \shp\label{eq:HeffFirstOrder}.
\end{align}
Here $\check{\chi} = {\xi}_g\ketbra{g}{g} + {\xi}_h\ketbra{h}{h}$, ${\xi}_{g/h} = \chi_{g/h} + \omega_r - \omega_d$, $\delta_q = \delta_h - \delta_g$.  Note that ${\xi}_{g/h}$ is redefined from Eq.~\eqref{eqn:Heff} by absorbing the resonator detuning for compactness, which introduces an overall energy shift $\delta_h+\delta_g$ that can be safely discarded in the subsequent analysis. After including resonator relaxation, we will first determine the steady-state solution of the zeroth-order Hamiltonian, and then treat the two-photon resonance term as a perturbation to obtain the final steady state and the corresponding driven–dissipative dynamics.

\subsection{Zeroth-Order Steady States without Two-Photon Resonance}
\label{sec:SteadyStateSolution}
We first solve the Lindblad master equation for the zeroth-order Hamiltonian $\hat{H}^{(0)}$:
\begin{align}
    \mathcal{L}_0\rhoh = -i[\hat{H}^{(0)}, \rhoh] + \kappa \mathcal{D}[\ah] \rhoh 
    = -i(\hat{H}_{\rm NH}\rhoh - \rhoh\hat{H}^{\dagger}_{\rm NH}) + \kappa\ah \rhoh \ad,
\end{align}
where the non-Hermitian Hamiltonian is defined as $\hat{H}_{\rm NH} = \hat{H}^{(0)} - \frac{i\kappa}{2} \ad\ah$.

To eliminate the drive term in $\hat{H}_{\rm NH}$, we move to a conditional displaced frame by defining
\begin{align}
&\hat{D}_c = \exp(\alphah\ad - \alphad\ah), \\
    &\alphah \equiv \alpha_h\ketbra{h}{h} + \alpha_g \ketbra{g}{g},
\end{align}
with the commutation relations $[\alphad, \alphah] = [\shz, \alphah] = 0$.  Under the conditional displacement transformation,
\begin{align}
    \hat{D}_c^\dagger \ah \hat{D}_c = \ah + \alphah, \quad \hat{D}_c^\dagger \shz \hat{D}_c = \shz.
\end{align}
The master equation in the transformed frame becomes
\begin{align}
\dot{\check{\rho}}= \check{\mathcal{L}}_0\check{\rho} = -i(\check{H}_{\rm NH}\check{\rho} -  \check{\rho}\check{H}_{\rm NH}^\dagger) + \kappa (\ah + \alphah) \check{\rho} (\ad + \alphad),
\end{align}
where $\check{\rho} = \hat{D}_c^\dagger \rhoh \hat{D}_c$ and the transformed non-Hermitian Hamiltonian is
\begin{align}
    \check{H}_{\rm NH} &= \hat{D}_c^\dagger \hat{H}_{\rm NH} \hat{D}_c\notag\\
    & = (\check{\chi} - \frac{i\kappa}{2})\ad\ah + \left((\check{\chi} - \frac{i\kappa}{2})\ad \alphah + \varepsilon_d \ad\right)  + \left((\check{\chi} - \frac{i\kappa}{2})\ah \alphad + \varepsilon_d\ah\right)+ (\check{\chi} - \frac{i\kappa}{2})\alphad\alphah + \frac{1}{2}\delta_q\shz + \varepsilon_d \alphad + \varepsilon_d \alphah.
\end{align}

Choosing $\alpha_g=\frac{-\varepsilon_d}{{\xi}_g - i\kappa/2}$ and $\alpha_h=\frac{-\varepsilon_d}{{\xi}_h - i\kappa/2}$ such that $\alphah = -\frac{\varepsilon_d }{\check{\chi} - i\kappa/2}$ eliminates the linear drive term from the transformed non-Hermitian Hamiltonian, yielding
\begin{align}
    \hat{H}_{\rm NH}  = (\check{\chi} - \frac{i\kappa}{2})\ad\ah - i\kappa\alphad \ah - \frac{i\kappa}{2}\alphad\alphah +\frac{1}{2}{\omega^{d}_q}\shz + {C},
\end{align}
where

\begin{align}
    \omega^{d}_q
    &=
    \delta_q+{\xi}_g|\alpha_g|^2-{\xi}_h|\alpha_h|^2,\\
    C
    &=
    \frac{1}{2}\left({\xi}_h|\alpha_h|^2+{\xi}_g|\alpha_g|^2\right).
\end{align}
The zeroth-order master equation in the transformed frame now reads
\begin{align}
    \check{\mathcal{L}}_0\check{\rho} = -i[\check{\chi}\ad\ah + \frac{1}{2}\omega^{d}_q\shz, \rhot] + \kappa \mathcal{D}[\ah] \rhot + \kappa \ah[\rhot, \alphad] + \kappa [\alphah, \rhot]\ad + \kappa \mathcal{D}[\alphah] \rhot,
    \label{eqn:L0t}
\end{align}
which admits the steady-state solution
\begin{align}
    \rhot^{(0)}_{ss} = \left( \tilde{P}_g\ketbra{g}{g} + \tilde{P}_h\ketbra{h}{h} \right)\otimes \ketbra{0}{0},
    \label{eqn:rhot0ss}
\end{align}
with arbitrary $\tilde{P}_g,\tilde{P}_h \geq 0$ satisfying $\tilde{P}_g + \tilde{P}_h = 1$, showing a degenerate steady-state subspace.  Moving back to the non-displaced frame, the zeroth-order steady state takes the form
\begin{align}
    \hat{\rho}_{ss}^{(0)} = \hat{D}_c\check{\rho}_{ss}^{(0)}\hat{D}_c^\dagger =  \tilde{P}_g\ketbra{g}{g}\otimes\ketbra{\alpha_g}{\alpha_g} + \tilde{P}_h\ketbra{h}{h}\otimes\ketbra{\alpha_h}{\alpha_h},
\end{align}
where $\ket{\alpha_{g/h}}$ are coherent states with amplitudes $\alpha_{g/h}$.

\subsection{Unique Steady State and Relaxation Rates under Two-Photon Resonance}
\label{sec:SecondOrderSolution}
Building on the zeroth-order analysis in Sec.~\ref{sec:SteadyStateSolution}, $\check{\mathcal{L}}_0$ drives the system into the degenerate steady-state manifold on the fast timescale set by $1/\kappa$. We now include the two-photon transition term $H^{(1)}$ as a perturbation to the dynamics on slower timescales. Since the rapid zeroth-order relaxation keeps the state confined to this manifold, we take $\check{\rho}^{(0)}$ to be the steady-state solution of Sec.~\ref{sec:SteadyStateSolution} throughout the perturbative expansion,
\begin{equation}
    \check{\rho}^{(0)} = \left( \tilde{P}_g\ketbra{g}{g} + \tilde{P}_h\ketbra{h}{h} \right)\otimes \ketbra{0}{0}.
    \label{eqn:rhot0}
\end{equation}
The full Liouvillian in the transformed frame is
\begin{equation}
    \dot{\check{\rho}} = \check{\mathcal{L}}_0 \check{\rho} + \check{\mathcal{L}}_1 \check{\rho},
\end{equation}
where $\check{\mathcal{L}}_0$ is defined in Eq.~\eqref{eqn:L0t} and the first-order Liouvillian is given by
\begin{align}
    \check{\mathcal{L}}_1 \rhot &= -i[\check{H}^{(1)}, \rhot], \\
    \check{H}^{(1)} &= \hat{D}_c^\dagger \hat{H}^{(1)} \hat{D}_c = g_{\rm eff}(\hat{A}^{(2)})^\dagger \shm + \text{H.C.}, \\
    \hat{A}^{(k)} &\equiv \hat{D}^\dagger(\alpha_h) \ah^k \hat{D}(\alpha_g).
 \end{align}
The total density matrix can be expressed in the two-level qubit subspace as
\begin{align}
    \rhot = \rhot_{gg}\ketbra{g}{g} + \rhot_{hh}\ketbra{h}{h} + \rhot_{hg}\ketbra{h}{g} + \rhot_{gh}\ketbra{g}{h},
\end{align}
where each $\rhot_{ij}$ is an operator acting on the resonator Hilbert space, and $\rhot_{hg}\phantom{^\dagger} = \rhot_{gh}^\dagger$ ensures the Hermiticity of the density matrix.

To proceed, we carry out a perturbative expansion of the dynamics order by order. We start with the first-order correction by expanding the density matrix as $\rhot \approx \rhot^{(0)} + \rhot^{(1)}$,
where $\rhot^{(0)}$ is block-diagonal in the qubit basis and belongs to the degenerate steady-state manifold given in Eq.~\eqref{eqn:rhot0}. In this structure, the off-diagonal elements $\rhot_{hg}$ and $\rhot_{gh}$ vanish at the zeroth order and are generated only through the coupling induced by $\check{\mathcal{L}}_1$. Under the limit $\abs{g_{\rm{eff}}}\abs{\alpha_g}^2, \abs{g_{\rm{eff}}}\abs{\alpha_h}^2 \ll \kappa$, the zeroth-order relaxation driven by $\check{\mathcal{L}}_0$ occurs on a timescale much shorter than the two-photon transition, allowing us to adiabatically eliminate $\rhot_{hg}$ by setting $\dot{\rhot}_{hg} = \dot{\rhot}_{gh}^\dagger \approx 0$.

Setting $\dot{\rhot}_{hg} = 0$, the off-diagonal term adiabatically follows the block-diagonal populations, satisfying
\begin{equation}
\dot{\check{\rho}}^{(1)}_{hg}=\left(\check{\mathcal{L}}_0 {\check{\rho}}^{(1)}\right)_{hg} + \left(\check{\mathcal{L}}_1 \rhot^{(0)}\right)_{hg} = 0.
    \label{eqn:AdiabaticElimination}
\end{equation}
Equivalently, the first-order coherence block can be written formally as
\begin{align}
    \check{\rho}_{hg}^{(1)}
    =
    -\check{\mathcal{L}}_{0,hg}^{-1}
    \left[
    \left(\check{\mathcal{L}}_1 \rhot^{(0)}\right)_{hg}
    \right],
\end{align}
where $\check{\mathcal{L}}_{0,hg}$ denotes the zeroth-order Liouvillian acting in the $\ketbra{h}{g}$ coherence sector,
\begin{align}
    \check{\mathcal{L}}_{0,hg}(\cdot)
    \equiv
    \bra{h}\check{\mathcal{L}}_0\!\left[\ketbra{h}{g}\otimes(\cdot)\right]\ket{g}.
\end{align}

Next, we extend the perturbative expansion to second order, $\rhot= \rhot^{(0)} +\rhot^{(1)}+\rhot^{(2)}$.  The first-order correction describes how the off-diagonal elements adiabatically follow the diagonal populations. The second-order correction will capture the resulting slow population transfer between the qubit ground and excited states, which governs the non-equilibrium dynamics and determines the unique steady state. The second-order dynamics is given by
\begin{align}
    \dot{\rhot}^{(2)} = \check{\mathcal{L}}_0 (\rhot^{(2)}) + \check{\mathcal{L}}_1 (\rhot^{(1)}).
\end{align}

Assume limit $\abs{g_{\rm{eff}}}\abs{\alpha_g}^2, \abs{g_{\rm{eff}}}\abs{\alpha_h}^2 \ll \kappa$, the fast dynamics of $\check{\mathcal{L}}_0$ quickly relaxes the system into the subspace $\{ \ket{g,0}$, $\ket{h,0} \}$. Within this subspace, the second-order dynamics reads
\begin{align}
    \dot{\rhot}^{(2)}
    & \approx \sum_{i\in\{g,h\}} \Tr_r\!\left[
        \ketbra{i}{i}\check{\mathcal{L}}_1(\rhot^{(1)})\ketbra{i}{i}
      \right] \otimes \ketbra{0}{0} \notag\\
    &= -ig_{\rm eff} \,\Tr_r \!\left[
    (\hat{A}^{(2)})^\dagger\rhot_{hg} \ketbra{g}{g}
        - \rhot_{hg}(\hat{A}^{(2)})^\dagger \ketbra{h}{h}
      \right]\otimes\ketbra{0}{0} + \text{H.c.} \label{eqn:rhot2dot} \\
    &\equiv \left(\dot{\tilde P}_g \ketbra{g}{g} + \dot{\tilde P}_h \ketbra{h}{h}\right) \otimes \ketbra{0}{0},\notag
\end{align}
where $\Tr_r$ is the partial trace over the degrees of freedom of the resonator.

Inserting the solution of $\rhot_{hg}$ into Eq.~\eqref{eqn:rhot2dot}, 
we obtain the closed rate equations for the qubit populations:
\begin{align}
    \dot{\tilde{P}}_g &= -\gamma_g \tilde{P}_g + \gamma_h \tilde{P}_h, \quad 
    \dot{\tilde{P}}_h = -\dot{\tilde{P}}_g,
\end{align}
where the transition rates $\gamma_g$ and $\gamma_h$ are formally given by
\begin{align}
    \gamma_g
    &=
    -2\abs{g_{\rm eff}}^2
    \Re \left\{
    \operatorname{Tr}_r
    \left[
    \hat{A}^{(2)\dagger}
   \check{\mathcal{L}}_{0,hg}^{-1} \left(
    \hat{A}^{(2)}\ketbra{0}{0}\right)
    \right]\right\},
    \notag\\
    \gamma_h
    &=
    -2\abs{g_{\rm eff}}^2
    \Re
      \left\{\operatorname{Tr}_r
    \left[
   \hat{A}^{(2)\dagger}
  \check{\mathcal{L}}_{0,hg}^{-1}
    \left(\ketbra{0}{0}\hat{A}^{(2)}\right)
    \right]\right\}.
\end{align}

To evaluate these formal expressions explicitly, we expand the coherence block in the resonator Fock basis. Expanding $\check{\rho}_{hg}=\check{\rho}^{(1)}_{hg}$ in the resonator Fock basis,
$\check{\rho}_{hg} = \sum_{nm} \rho_{hg}^{nm}\ketbra{n}{m}$, we obtain
\begin{align}
    \check{\mathcal{L}}_0 \left(\check{\rho}_{hg}^{(1)}\shp \right)
    &= \sum_{nm} \ketbra{n}{m}\shp \left(c_{nm}^{00} \rho_{hg}^{nm} + c_{nm}^{10}\rho_{hg}^{n+1,m} + c_{nm}^{01}\rho_{hg}^{n, m+1} + c_{nm}^{11}\rho_{hg}^{n+1, m+1}\right),\notag\\
      \left(\check{\mathcal{L}}_1 \check{\rho}^{(0)}\right)_{hg} &= -ig_{\rm eff}^*\left(\tilde{P}_g \hat{A}^{(2)}\ketbra{0}{0} - \tilde{P}_h \ketbra{0}{0}\hat{A}^{(2)}\right) = -ig_{\rm eff}^* \sum_{nm} \left(\tilde{P}_g A^{(2)}_{n0}\delta_{0m} - \tilde{P}_h A^{(2)}_{0m}\delta_{n0}\right)\ketbra{n}{m},
        \label{eqn:L0L1expansion}
    \end{align}
    where the coefficients are given by
    \begin{align}
    c^{00}_{nm} &= -i(\xi_h n - \xi_g m + \omega^{d}_q) - \frac{\kappa}{2}(n + m) + \kappa(\alpha_g ^*\alpha_h - \frac{1}{2}(|\alpha_h|^2 + |\alpha_g|^2)) \notag\\
                &= \left(-i\xi_h - \frac{\kappa}{2}\right)n + \left(i\xi_g- \frac{\kappa}{2}\right)m - i\left(\xi_h - \xi_g\right)\alpha_g^*\alpha_h -i\delta_q,\\
    c^{10}_{nm} &= \kappa (\alpha_g^* - \alpha_h^*)\sqrt{n+1},\\
    c^{01}_{nm} &= -\kappa(\alpha_g - \alpha_h)\sqrt{m+1}, \\
    c^{11}_{nm} &= \kappa\sqrt{(n+1)(m+1)}.
\end{align}

Substituting the Fock-basis expansions of $\check{\mathcal{L}}_0(\rhot_{hg}^{(1)}\shp)$ 
and $(\check{\mathcal{L}}_1\rhot^{(0)})_{hg}$ from Eq.~\eqref{eqn:L0L1expansion} into Eq.~\eqref{eqn:AdiabaticElimination} yields
\begin{align}
    c_{nm}^{00} \rho_{hg}^{nm} + c_{nm}^{10}\rho_{hg}^{n+1,m} + c_{nm}^{01}\rho_{hg}^{n, m+1} + c_{nm}^{11}\rho_{hg}^{n+1, m+1} = ig_{\rm eff}^*(\tilde{P}_g A^{(2)}_{n0}\delta_{0m} - \tilde{P}_h A^{(2)}_{0m}\delta_{n0}).
\end{align}
Because the right-hand side is nonzero only when either $n=0$ or $m=0$, it follows that $\rho_{hg}^{nm}$ is nonzero only in these cases.  We can therefore write
\begin{align}
    \rho^{nm}_{hg} &= ig_{\rm eff}^*(\tilde{P}_g x^g_n\delta_{0m} - \tilde{P}_h x^{h}_m \delta_{n0}),    
    \label{eqn:rhonmhg}\\
    \rhot_{hg} &= ig_{\rm eff}^* \left(\sum_{n=0}^{\infty} \tilde{P}_g x_{n}^{g}\ketbra{n}{0} - \sum_{n=0}^{\infty} \tilde{P}_h x_{n}^{h} \ketbra{0}{n}\right),
    \label{eqn:rhohg}
\end{align} where $x_n^g$ and $x_n^h$ satisfy
\begin{align}
    c_{n0}^{00}x^g_n + c_{n0}^{10}x^g_{n+1} &= A^{(2)}_{n0},\\
    c_{0n}^{00}x^h_n + c_{0n}^{01}x^h_{n+1} &= A^{(2)}_{0n}.
\end{align}
Both recurrence relations can be cast into the generic form $b_{n}x_n + d_{n}x_{n+1} = y_n$, whose formal solution is
\begin{align}
    x_n = \sum_{\ell=n}^\infty (-1)^{\ell-n}\frac{y_{\ell}}{b_{\ell}}\left(\prod_{k=n}^{\ell-1}\frac{d_k}{b_k}\right).
\end{align}
Applying this solution to the two specific cases, we find
\begin{align}
    x_n^g &= \sum_{\ell=n}^{\infty}(-1)^{\ell-n} \frac{A^{(2)}_{\ell0}}{c^{00}_{\ell0}}\left(\prod_{k=n}^{\ell-1}\frac{c^{10}_{k0}}{c^{00}_{k0}}\right),\\
    x_n^h &= \sum_{\ell=n}^{\infty}(-1)^{\ell-n} \frac{A^{(2)}_{0\ell}}{c^{00}_{0\ell}}\left(\prod_{k=n}^{\ell-1}\frac{c^{01}_{0k}}{c^{00}_{0k}}\right).
\end{align}

Using these explicit expressions, the transition rates can be written as
\begin{align}
    \gamma_g &= -2\abs{g_{\rm eff}}^2 \,\Re\!\left(\sum_{n=0}^{\infty} \left(A^{(2)}_{n0}\right)^* x_n^g\right), \label{eq:Gamma_g}\\
    \gamma_h &= -2|g_{\rm eff}|^2 \,\Re\!\left(\sum_{n=0}^{\infty} \left(A^{(2)}_{0n}\right)^* x_n^h\right). \label{eq:Gamma_e}
\end{align}

Imposing the normalization condition $\tilde{P}_g + \tilde{P}_h = 1$, the dynamics reduces to $\dot{\tilde{P}}_g = -\gamma (\tilde{P}_g -\tilde{P}_g^{\rm{ss}})$, where
\begin{align}
    \quad \gamma \equiv \gamma_g + \gamma_h, 
    \quad \tilde{P}_g^{\rm{ss}} = \frac{\gamma_h}{\gamma},  \quad \tilde{P}_h^{\rm{ss}} = \frac{\gamma_g}{\gamma}.
\end{align}
The rate equations remain unchanged when transformed back to the original frame, and describe the relaxation of the qubit populations toward a unique steady-state distribution $\tilde{P}_g^{\rm{ss}}$ and $\tilde{P}_h^{\rm{ss}}$.

\subsection{Multi-Photon Resonances: Generalized Dynamics and Rate Equations}

The perturbative approach in Sec.~\ref{sec:SecondOrderSolution} can be generalized to multi-photon transitions.
We consider a $k$-photon resonance between a lower state $\ket{\mu}$ and an excited qubit state $\ket{\nu}$, with $\omega_\nu - \omega_\mu \approx k\omega_r$.
The dominant dynamics may be captured by the effective Hamiltonian, up to higher-order corrections to the resonator energy,
\begin{align}
    \hat{H}_{\rm eff} &= \hat{H}^{(0)} + \hat{H}^{(1)},\\
    \hat{H}^{(0)} &= \check{\chi}\,\ad\ah + \frac{1}{2}\delta_q\shz + \varepsilon_d(\ad + \ah) + \rm const,\\
    \hat{H}^{(1)} &= g_{\rm eff}^{(k)} (\ad)^k \shm + \big(g_{\rm eff}^{(k)}\big)^* \ah^k \shp,
\end{align}
where $\shz$, $\shm$, and $\shp$ are defined in the effective two-level subspace spanned by $\ket{\mu}$ and $\ket{\nu}$.

The steady state of $\hat{H}^{(0)}$ is the same as in Sec.~\ref{sec:SteadyStateSolution}.
Applying the same adiabatic elimination procedure as in Sec.~\ref{sec:SecondOrderSolution}, under the condition $\abs{g_{\rm{eff}}}\abs{\alpha_\mu}^k, \abs{g_{\rm{eff}}}\abs{\alpha_\nu}^k \ll \kappa$, where $\alpha_{\mu/\nu} = \frac{-\varepsilon_d}{\xi_{\mu/\nu} - i\kappa/2}$, we obtain the $k$-photon transition rates
\begin{align}
    \gamma_\mu^{(k)} &= -2\big|g_{\rm eff}^{(k)}\big|^2 \,\Re\!\left( \sum_{n=0}^\infty \big(A^{(k)}_{n0}\big)^*\, x_n^\mu \right),\\
    \gamma_\nu^{(k)} &= -2\big|g_{\rm eff}^{(k)}\big|^2 \,\Re\!\left( \sum_{n=0}^\infty \big(A^{(k)}_{0n}\big)^*\, x_n^\nu \right),
\end{align}
where
\begin{align}
    x_n^\mu &= \sum_{m=n}^\infty (-1)^{m-n} \frac{A^{(k)}_{m0}}{c^{00}_{m0}}
    \left( \prod_{\ell=n}^{m-1} \frac{c^{10}_{\ell 0}}{c^{00}_{\ell 0}} \right),\\
    x_n^\nu &= \sum_{m=n}^\infty (-1)^{m-n} \frac{A^{(k)}_{0m}}{c^{00}_{0m}}
    \left( \prod_{\ell=n}^{m-1} \frac{c^{01}_{0\ell}}{c^{00}_{0\ell}} \right).
\end{align}
The coefficients $A^{(k)}_{nm}$ and $c^{ab}_{nm}$ follow expressions analogous to those in Sec.~\ref{sec:SecondOrderSolution} under the straightforward $k$-photon generalization, thereby determining the relaxation dynamics between $\ket{\mu}$ and $\ket{\nu}$.

\section{Transition Rates in Different Physical Regimes}
\subsection{Zero-drive Limit}
In the zero-drive limit, $\varepsilon_d=0$, one has $\alpha_g=\alpha_h=0$, so that $\hat A^{(2)}=\hat a^2$. The formal rate expressions then reduce to
\begin{align}
    \gamma_g
    &=
    -2|g_{\rm eff}|^2
    \Re\!\left\{
    \operatorname{Tr}_r\!\left[
    \hat a^{\dagger 2}
    \mathcal{L}_{0,hg}^{-1}\!\left(
    \hat a^2\ketbra{0}{0}\right)
    \right]\right\},\\
    \gamma_h
    &=
    -2|g_{\rm eff}|^2
    \Re\!\left\{
    \operatorname{Tr}_r\!\left[
    \hat a^{\dagger 2}
    \mathcal{L}_{0,hg}^{-1}\!\left(
    \ketbra{0}{0}\hat a^2\right)
    \right]\right\}.
\end{align}
Since $ \hat a^2\ketbra{0}{0} = 0$
we immediately obtain $\gamma_g=0$. For the downward transition, $\ketbra{0}{0}\hat a^2 = \sqrt2\,\ketbra{0}{2}$,
we find
\begin{align}   \mathcal{L}_{0,hg}\!\left(\ketbra{0}{2}\right)
    &=
    -i\left[
    \left(\frac{\delta_q}{2}+\xi_h\ad\ah\right)\ketbra{0}{2}
    -
    \ketbra{0}{2}\left(-\frac{\delta_q}{2}+\xi_g\ad\ah\right)
    \right]
    +\kappa \mathcal D[\hat a]\!\left(\ketbra{0}{2}\right) \notag\\
    &=
    \left[-\kappa-i(\delta_q-2\xi_g)\right]\ketbra{0}{2},\notag\\
\mathcal{L}_{0,hg}^{-1}\!\left(\ketbra{0}{2}\right)
    &=
    -\frac{1}{\kappa+i(\delta_q-2\xi_g)}\ketbra{0}{2}.
\end{align}
Substituting this into the expression for $\gamma_h$, we obtain
\begin{align}
    \gamma_h
    =
    4|g_{\rm eff}|^2
    \frac{\kappa}{(\delta_q-2\xi_g)^2+\kappa^2}.
    \label{eqn:ZeroDriveLimit}
\end{align}
This result admits a simple Fermi's golden rule interpretation as a two-photon Purcell process: the transition $\ket{h,0}\to\ket{g,2}$ is driven by the matrix element $\sqrt{2}\,g_{\rm eff}$, while the two-photon state $\ket{g,2}$ is broadened by resonator dissipation at the total photon-loss rate $2\kappa$, yielding the Lorentzian denominator in Eq.~\eqref{eqn:ZeroDriveLimit}.

\subsection{Gaussian Approximation in the Resolved Regime}

\begin{figure}[H]
    \centering
\includegraphics[width=0.7\linewidth]{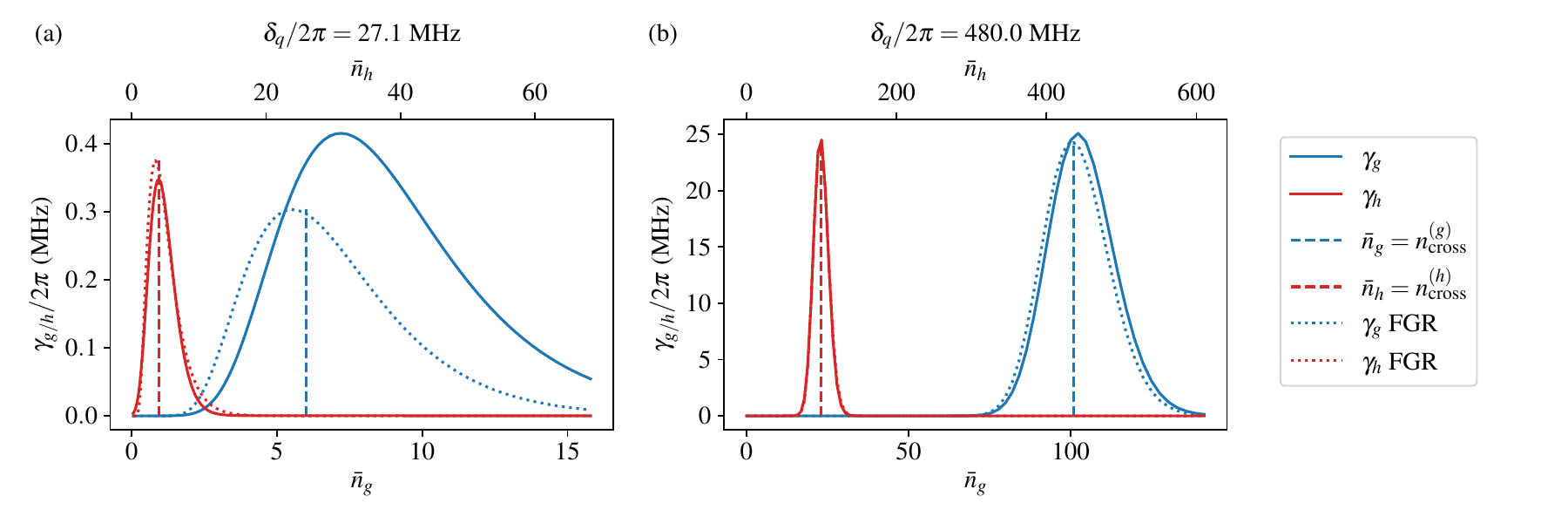}
\includegraphics[width=0.7\linewidth]{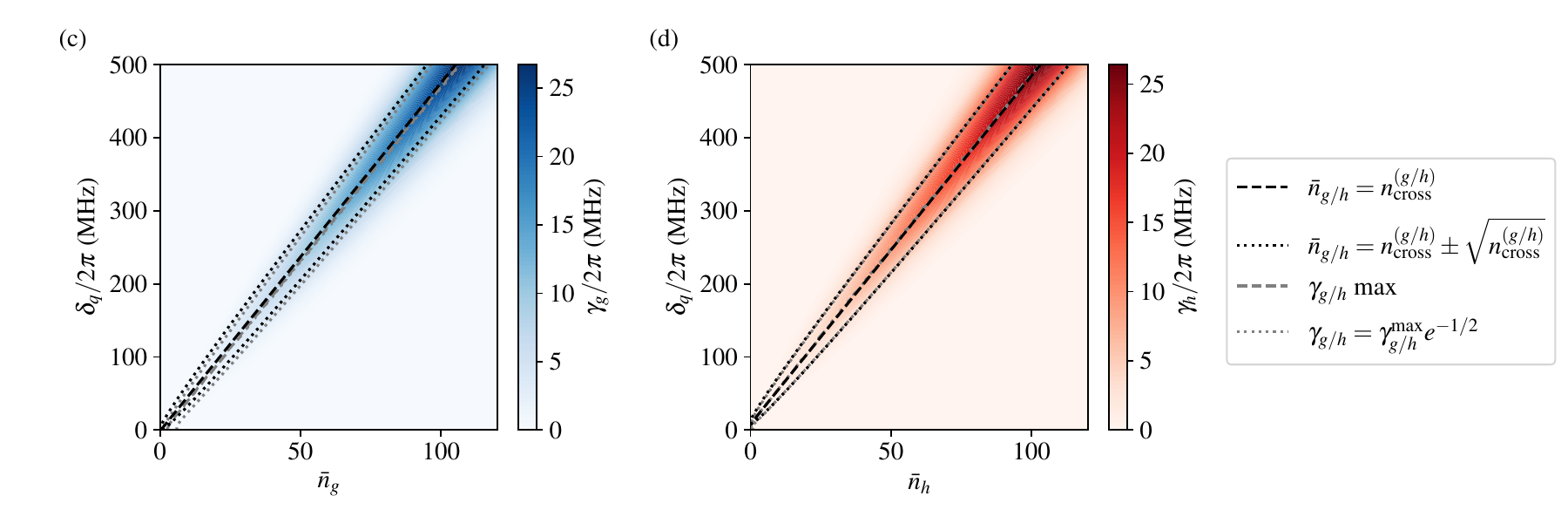}
    \caption{
    Comparison between the exact transition rates and the Gaussian approximations in the resolved regime. 
    (a,b) Exact rates $\gamma_{g/h}$ obtained from Eqs.~\eqref{eq:Gamma_g} and \eqref{eq:Gamma_e}, compared with the Gaussian approximation derived in Eq.~(C16) of the main text, for (a) $\delta_q/2\pi = 27.1\,\mathrm{MHz}$, corresponding to the main-text parameters, and (b) $\delta_q/2\pi = 480\,\mathrm{MHz}$. 
    (c,d) Exact rates plotted as functions of $n_{g/h}$, showing maxima near $n_{g/h}=n_{\rm cross}^{(g/h)}$. The $1\sigma$ contours, defined by $\gamma_{g/h}=\gamma_{g/h}^{\rm max}e^{-1/2}$, occur near $n_{g/h}=n_{\rm cross}^{(g/h)}\pm\sqrt{n_{\rm cross}^{(g/h)}}$, consistent with the Gaussian approximation.
    }
\label{fig:RelaxationRateVsDelta_q}
\end{figure}

The branch-resolved Gaussian approximation is derived in Appendix~C of the main text. Here we compare it with the exact rates obtained from Eqs.~\eqref{eq:Gamma_g} and \eqref{eq:Gamma_e}. As shown in Fig.~\ref{fig:RelaxationRateVsDelta_q}(a,b), the Gaussian approximation captures the overall resolved-regime profile and becomes more accurate at larger $\delta_q$, where the relevant crossing photon numbers are larger and the Poissonian photon-number distributions are better approximated by Gaussian envelopes. Figures~\ref{fig:RelaxationRateVsDelta_q}(c,d) further show that the exact rates are maximized near $n_{g/h}=n_{\rm cross}^{(g/h)}$, with the $1\sigma$ contours approximately given by
\begin{align}
    n_{g/h}
    =
    n_{\rm cross}^{(g/h)}
    \pm
    \sqrt{n_{\rm cross}^{(g/h)}}.
\end{align}

\subsection{Lorentzian Approximation in the Unresolved Regime}
\label{sec:UnresLorentzian}

We consider the unresolved regime near the transition point, characterized by the small distinguishability
\begin{align}
    D
    \equiv
    \abs{\alpha_g-\alpha_h}^2
    \ll 1 .
\end{align}
In this limit, the two qubit-conditioned resonator states strongly overlap and the resonator response can be approximated by a common coherent amplitude. Starting from
\begin{align}
    \alpha_{g/h}
    =
    -\frac{\varepsilon_d}{\delta_a+\chi_{g/h}-i\kappa/2},
\end{align}
and assuming
\begin{align}
    \abs{\chi_{g/h}}
    \ll
    \abs{\delta_a-i\kappa/2},
\end{align}
we obtain
\begin{align}
    \alpha_{g/h}
    \approx
    \alpha_c
    \left[
        1+
        \mathcal{O}\!\left(
        \frac{\chi_{g/h}}{\delta_a-i\kappa/2}
        \right)
    \right],
    \qquad
    \alpha_c
    =
    -\frac{\varepsilon_d}{\delta_a-i\kappa/2},
    \qquad
    n_c=\abs{\alpha_c}^2, \quad D \approx n_c\frac{{(\chi_g - \chi_h)}^2}{|\delta_a-i\kappa/2|^2}.
\end{align}

To extract the leading unresolved contribution, we expand the displaced two-photon matrix elements in the Fock basis,
\begin{align}
    A^{(2)}_{n0}
    &=
    \frac{1}{\sqrt{n!}}
    (\alpha_g-\alpha_h)^n
    \alpha_g^2
    \langle\alpha_h|\alpha_g\rangle,
    \\
    A^{(2)}_{0n}
    &=
    \frac{1}{\sqrt{n!}}
    (\alpha_h^*-\alpha_g^*)^n
    \alpha_g^2
    \langle\alpha_h|\alpha_g\rangle
    f_n,
\end{align}
where
\begin{align}
    f_n
    =
    1
    +
    \frac{2n}{\alpha_g(\alpha_h^*-\alpha_g^*)}
    +
    \frac{n(n-1)}
    {\alpha_g^2(\alpha_h^*-\alpha_g^*)^2}.
\end{align}
Substituting these expressions into Eqs.~\eqref{eq:Gamma_g} and \eqref{eq:Gamma_e}, the rates can be reorganized as
\begin{align}
    \gamma_g
    &=
    2\abs{g_{\rm eff}}^2\abs{\alpha_g}^4
    \sum_{n=0}^{\infty}
    \frac{D^n}{n!}e^{-D}
    \sum_{\ell=0}^{\infty}
    \kappa^\ell D^\ell
    \Re\left[
        \prod_{k=n}^{n+\ell}
        \frac{-1}{c_{k0}^{00}}
    \right],
    \label{eq:UnresGammaGRefactor}
    \\
    \gamma_h
    &=
    2\abs{g_{\rm eff}}^2\abs{\alpha_g}^4
    \sum_{n=0}^{\infty}
    \frac{D^n}{n!}e^{-D}
    \sum_{\ell=0}^{\infty}
    \kappa^\ell D^\ell
    \Re\left[
        f_n^*f_{n+\ell}
        \prod_{k=n}^{n+\ell}
        \frac{-1}{c_{0k}^{00}}
    \right].
    \label{eq:UnresGammaHRefactor}
\end{align}
For $D\ll1$, the leading coherent-field contribution comes from the $(n,\ell)=(0,0)$ term. For the upward transition,
\begin{align}
    \gamma_g
    &\approx
    2\abs{g_{\rm eff}}^2\abs{\alpha_g}^4
    \Re\left[
    \frac{-1}{c_{00}^{00}}
    \right]
    \notag\\
    &=
    2\abs{g_{\rm eff}}^2\abs{\alpha_g}^4
    \Re\left[
    \frac{1}{
    i\left[
    \delta_q-(\chi_g-\chi_h)\alpha_g^*\alpha_h
    \right]
    }
    \right]
    \notag\\
    &\approx
    2\abs{g_{\rm eff}}^2 n_c^2
    \Re\left[
    \frac{1}{
    i\left[
    \delta_q-(\chi_g-\chi_h)n_c
    \right]
    +\Gamma_\phi}
    \right],
    \label{eq:UnresGammaGIntermediate}
\end{align}
where
\begin{align}
    \Gamma_\phi
    =
    \frac{\kappa D}{2}
    \label{eq:UnresGammaPhi}
\end{align}
is the measurement-induced dephasing rate generated by the residual distinguishability between the two conditional resonator states.

For the downward transition,
\begin{align}
    \frac{\gamma_h}{\gamma_g} &\approx \frac{
        \Re{|f_0|^2\frac{-1}{c_{00}^{00}} + \kappa D f_0^*f_1\frac{1}{c_{00}^{00}c_{01}^{00}} 
        + D|f_1|^2 \frac{-1}{c_{01}^{00}}
        + \frac{1}{2}D^2|f_2|^2 \frac{-1}{c_{02}^{00}}
    }}{\Re{\frac{-1}{c_{00}^{00}}}} \\
    &\approx 1 + \frac{4(n^{\rm (unres)}_{\rm cross} - n_c)}{n_c} + (n^{\rm (unres)}_{\rm cross} - n_c)^2\left(\frac{4}{n_c^2} + \frac{1}{n_c^3}\right),
 \label{eqn:LargeKappaProbabilityRatio}
\end{align}
which approaches 1 near the transition as $n^{\rm (unres)}_{\rm cross} \approx n_c$.

\begin{figure}
    \centering
    \includegraphics[width=0.9\linewidth]{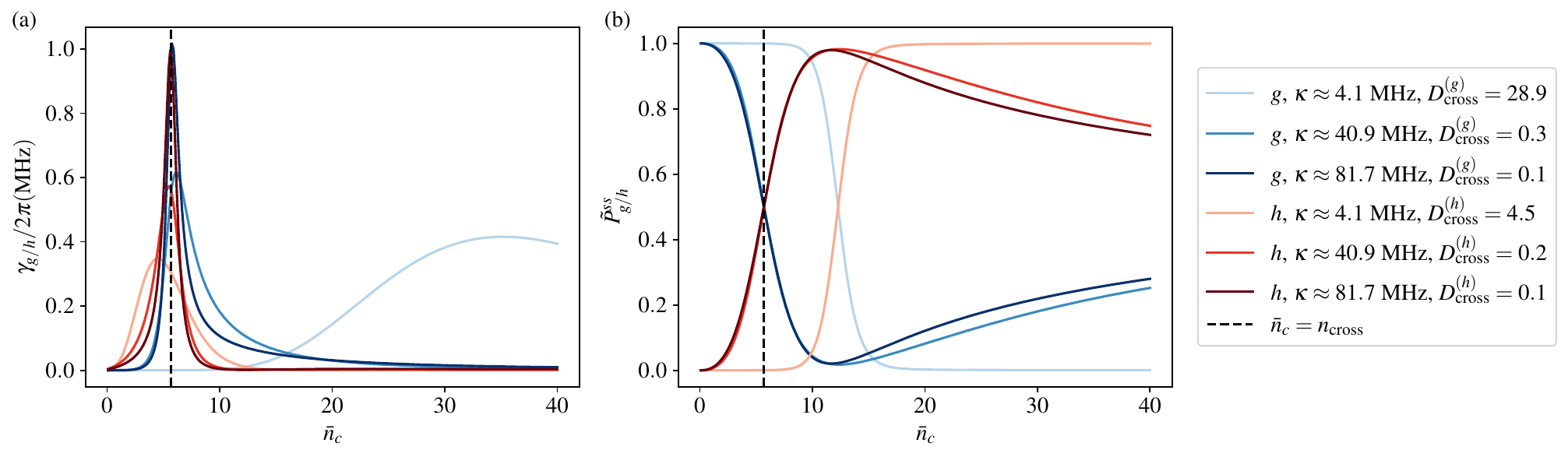}
    \caption{
    (a) Relaxation rates $\gamma_g$ and $\gamma_h$ and
    (b) steady-state populations $\tilde{P}_{g/h}^{\rm ss}$ in the unresolved regime.
    Different curves correspond to different resonator decay rates $\kappa$, while the other reduced-model parameters are fixed to the values in Table~II of the main text.
    The color bar labels each curve by $\kappa$ and by the distinguishability $D$ evaluated near the corresponding numerical transition point.
    The rate profiles approach a common Lorentzian-like peak around $n_c=n_{\rm cross}^{\rm unres}$, consistent with the leading unresolved approximation in Eq.~\eqref{eq:UnresLorentzian}.
    }
    \label{fig:UnresLorentzianRates}
\end{figure}

Taking the real part gives the leading unresolved Lorentzian rate
\begin{align}
    \gamma_g
    \approx
    \gamma_h
    \approx
    2\abs{g_{\rm eff}}^2 n_c^2
    \frac{\Gamma_\phi}
    {
    \left[
        \delta_q-(\chi_g-\chi_h)n_c
    \right]^2+\Gamma_\phi^2
    },
    \label{eq:UnresLorentzian}
\end{align}
centered at $n_c=n_{\rm cross}^{\rm unres}$, where
$n_{\rm cross}^{\rm unres}\equiv\delta_q/(\chi_g-\chi_h)$. This leading approximation agrees with the numerical rates shown in Fig.~\ref{fig:UnresLorentzianRates}, where the transition profiles become localized around $n_{\rm cross}^{\rm unres}$ in the unresolved regime.

\section{Numerical Simulations Setup}
In this section, we describe the simulation setup and analysis methods used for the numerical results in the main text. All numerical simulations are implemented in QuTiP~\cite{Lambert2026}. In the main text, we simulate the system with photon truncation number $n^{\rm max} = 100$, and qubit truncation number $j^{\rm max} = 4$. For the reduced model, we use a qubit subspace spanned by $\{\ket{g} = \ket{0}, \ket{h} = \ket{3}\}$, corresponding to the two-photon resonance $\omega_h - \omega_g \approx 2\omega_r$.

\subsection{Model Comparisons}
\label{sec:Model}

We compare dynamics from the full quantum model~\cite{dumas2024} and from a semiclassical approximation including measurement backaction~\cite{Wang2026}. In our simulations, this version of the semiclassical model shows closer agreement with the full quantum results than the backaction-free version.

For the full quantum model, the evolution follows the master equation, Eq.~\eqref{eqn:FullMasterEquation}.
In the semiclassical model, the resonator operators are replaced by their expectation values, $\alpha(t) = \langle \ah \rangle = \Tr\!\left(\ah\,\rhoh\right)$, leading to the equation of motion
\begin{equation}
    \dot{\alpha} = -i\omega_r\alpha + g\langle\hat{n}\rangle - 2\varepsilon_d\sin(\omega_d t) -\frac{\kappa}{2}\alpha.
    \label{eqn:SemiclassicalResonatorEOM}
\end{equation}
Substituting $\ah \to \alpha$ and $\ad \to \alpha^*$ in $\hat{H}_{\rm qr}$ yields a time-dependent qubit Hamiltonian,
\begin{align}
    \hat{H}_{\rm sc}(t) &= \hat{H}_q + 2g\,\hat{n}\,\Im\{\alpha(t)\}, \\
    i\frac{\partial}{\partial t}\ket{\psi_{\rm sc}(t)} &= \hat{H}_{\rm sc}(t)\ket{\psi_{\rm sc}(t)}.
    \label{eqn:SemiclassicalSchrodinger}
\end{align}

\subsection{Time Evolution Simulation Settings}
For each model, we apply a different numerical method for the time evolution simulation. 
For the full model, whose large Hilbert space and fast dynamics make direct integration expensive, we use the QuTiP Monte Carlo solver to solve the full master equation [Eq.~\eqref{eqn:FullMasterEquation}] (using 200 trajectories for $\varepsilon_d\leq {\rm 12MHz}$, and 1400 trajectories for $\varepsilon_d > {\rm 12MHz}$.) 
For the reduced model, we use the QuTiP master equation solver to integrate the master equation of the reduced quantum model [Eq.~\eqref{eqn:effMasterEquation}]. 
For the semiclassical model, we use a custom ODE solver to evolve the coupled nonlinear equations [Eqs.~\eqref{eqn:SemiclassicalResonatorEOM} and \eqref{eqn:SemiclassicalSchrodinger}]. 
A summary of the solvers, time steps, and truncations is given in Table~\ref{tab:SimulationParameters}.

Different initial states are used for the quantum and semiclassical models, reflecting their distinct treatments of the resonator mode. To ensure consistency between the quantum models, we initialize the reduced quantum model in the state $\ket{g,0}$ of its SW–transformed frame, while in the full model we use $\ket{\widetilde{g,0}} = e^{-\hat{S}}\ket{g,0}$, which corresponds to the same physical ground state expressed in the laboratory frame. For the semiclassical model, the initial qubit state is $\ket{g}$ with the resonator coherent-state amplitude set to $\alpha = 0$.

\renewcommand{\arraystretch}{1.15}
\begin{table}[htbp]
\centering
\caption{Numerical simulation settings for different models.}
\label{tab:SimulationParameters}
\begin{tabular}{l c c c}
\hline
Model & Full quantum model & Reduced quantum model & Semiclassical model \\
\hline
Solver & QuTiP Monte Carlo & QuTiP master equation & Custom ODE \\
Trajectories &
$200$ for $\varepsilon_d \le 12~\mathrm{MHz}$;
$1400$ for $\varepsilon_d > 12~\mathrm{MHz}$ &
--- &
--- \\
Time step (ns) & 2.5 & 2.5 & 2.5 \\
$n^{\rm max}$ & 100 & 100 & --- \\
$j^{\rm max}$ & 4 & --- & 4 \\
\hline
\end{tabular}
\end{table}
\renewcommand{\arraystretch}{1.0}

\subsection{Numerical Extraction of Relaxation Rates}
To extract the relaxation rates in Fig.~2(a) of the main text, we first obtain the steady-state solution of the master equation [Eq.~\eqref{eqn:effMasterEquation}] using the steady-state solver in QuTiP. This yields the steady-state dressed-state populations $\tilde{P}^{\rm ss}_{g/h}$. We then perform a time-evolution simulation of the reduced model to obtain $\tilde{P}_{g/h}(t)$ and fit the results to the exponential form
\begin{align}
    \tilde{P}_{g/h}(t) = \tilde{P}^{\rm ss}_{g/h} + C_{g/h} e^{-\gamma t},
    \label{eqn:DressedPopulationEvolution}
\end{align}
where $C_{g/h}$ is a time-independent constant determined by the initial conditions.
To avoid transient bias, only data with $t > t_{\rm min} \approx 10/\kappa$ are included. For improved accuracy, we initialize the system in $\ket{h,0}$ when $\varepsilon_d/2\pi \leq 7\,{\rm MHz}$ and in $\ket{g,0}$ otherwise, so that the fitted coefficient $C_{g/h}$ is sufficiently large for accurate extraction.

\subsection{Estimation of Dressed-State Populations}
Given the analytically obtained relaxation rate $\gamma$ and steady-state populations $\tilde{P}^{\rm ss}_{g/h}$ from Sec.~\ref{sec:SecondOrderSolution}, the dressed-state populations at any time $t$ can be estimated using Eq.~\eqref{eqn:DressedPopulationEvolution}. For the initial state $\ket{g,0}$, which gives $\tilde{P}_g(0)=1$,
\begin{align}
    \tilde{P}_{g/h}(t) \approx \tilde{P}^{\rm ss}_{g/h} + \big[1 - \tilde{P}^{\rm ss}_{g/h}\big] e^{-\gamma t}.
\end{align}
This expression is used to generate the curves shown in Fig.~4(d) of the main text.

\bibliographystyle{apsrev4-2}
\bibliography{MIST}